\documentclass[fleqn,usenatbib]{mnras}

\usepackage[T1]{fontenc}
\usepackage{ae,aecompl}

\usepackage{etoolbox}

\usepackage{graphicx}	
\usepackage{amsmath}	
\DeclareMathOperator{\sech}{sech}
\usepackage{amssymb}	
\usepackage[figure,figure*]{hypcap}
\usepackage{float}

\newcommand{\comm}[1]{}
\newcommand{\chisquare}{\ensuremath{\chi^{2}}}
\defcitealias{2019MNRAS.490.1539R}{R19}

\title[The structure of the Milky Way]{The structure of the Milky Way based on unWISE\footnote{Unblurred coadds of the Wide-field Infrared Survey Explorer imaging} 3.4~$\mu$m integrated photometry}

\author[A. V. Mosenkov et al.]{
Aleksandr V. Mosenkov,$^{1,2}$\thanks{E-mail: aleksandr\_mosenkov@byu.edu}
Sergey S. Savchenko,$^{3,4,2}$
Anton A. Smirnov,$^{2,3}$
\newauthor
and Peter Camps$^{5}$
\\
$^{1}$Department of Physics and Astronomy, N283 ESC, Brigham Young University, Provo, UT 84602, USA\\
$^{2}$Central Astronomical Observatory at Pulkovo of RAS, Pulkovskoye Chaussee 65/1, 196140 St. Petersburg, Russia\\
$^{3}$St.Petersburg State University, 7/9 Universitetskaya nab., St.Petersburg, 199034, Russia\\
$^{4}$Special Astrophysical Observatory, Russian Academy of Sciences, 369167 Nizhnij Arkhyz, Russia\\
$^{5}$Sterrenkundig Observatorium, Universiteit Gent, Krijgslaan 281, B-9000 Gent, Belgium\\\\
}

\date{Accepted XXX. Received YYY; in original form ZZZ}

\pubyear{2020}

\begin{document}
\label{firstpage}
\pagerange{\pageref{firstpage}--\pageref{lastpage}}
\maketitle

\begin{abstract}
We present a detailed analysis of the Galaxy structure using an unWISE wide-field image at $3.4\mu$m. We perform a 3D photometric decomposition of the Milky Way taking into account i) the projection of the Galaxy on the celestial sphere and ii) that the observer is located within the Galaxy at the solar radius. We consider a large set of photometric models starting with a pure disc model and ending with a complex model which consists of thin and thick discs plus a boxy-peanut-shaped bulge. In our final model, we incorporate many observed features of the Milky Way, such as the disc flaring and warping, several over-densities in the plane, and the dust extinction. The model of the bulge with the corresponding X-shape structure is obtained from N-body simulations of a Milky Way-like galaxy. This allows us to retrieve the parameters of the aforementioned stellar components, estimate their contribution to the total Galaxy luminosity, and constrain the position angle of the bar. The mass of the thick disc in our models is estimated to be 0.4-1.3 of that for the thin disc. The results of our decomposition can be directly compared to those obtained for external galaxies via multicomponent photometric decomposition.
\end{abstract}




\section{Introduction}
\label{sec:intro}

Our Galaxy represents a classical barred spiral galaxy, one among many others observed in the Local Universe. The location of the Earth within the Galactic disc makes it hard to trace the entire structure of the Milky Way, including the spiral arms, -- this is unavoidable if we investigate an edge-on galaxy. However, the advantage of this orientation is that we can explore the vertical structure of the disc and the galaxy's central components, the bulge and the bar. Multiple detailed studies of the Galactic structure (see e.g. a review by \citealt{2016ARA&A..54..529B} and references therein), using results of many space and ground-based surveys and the ongoing Gaia astrometric mission \citep{2016A&A...595A...1G}, underlie our understanding of the physical processes which led to its current view with its observed characteristics.

Based on global properties (its colour, star formation rate, and stellar mass), the Milky Way can be attributed to the sparsely populated ``green valley'' region in the galaxy color-magnitude diagram \citep{2015ApJ...809...96L}. Its place on some galaxy scaling relations (for example, on the Tully-Fisher relation, \citealt{1977A&A....54..661T}) is not typical for the majority of spiral galaxies \citep{2007ApJ...662..322H}, thus making it a somewhat unusual spiral.

Although there is an enormous number of studies dedicated to a comprehensive analysis of the Milky Way structural properties, both from observations (see e.g.  \citealt{1980ApJS...44...73B,1983MNRAS.202.1025G,1986ARA&A..24..577B,1995ApJ...445..716D,2001ApJ...553..184C,2008ApJ...673..864J,2014ApJ...783..130R,2020A&A...637A..96C}) and simulations of Milky Way-like galaxies (see e.g.  \citealt{2003ApJ...591..499A,2003A&A...409..523R,2017MNRAS.467.2430M,2020MNRAS.491.3461B,2020arXiv200606008A}), the estimates of its structural parameters are in poor agreement (see e.g. \citealt{2011ARA&A..49..301V,2016ApJ...831...71L,2016ARA&A..54..529B}). This can be explained by different reasons: various approaches used (e.g. star counts, \citealt{1980ApJS...44...73B,1986ARA&A..24..577B}; integrated light observations, \citealt{1998ApJ...492..495F,2016AJ....152...14N,2017MNRAS.471.3988C}), different data exploited (e.g., the Two Micron All Sky Survey, 2MASS, \citealt{2006AJ....131.1163S}; the Sloan Digital Sky Survey, SDSS, \citealt{2000AJ....120.1579Y}; \textit{Gaia}, \citealt{2016A&A...595A...1G}, etc.), wavelengths used (X-ray, \citealt{2012ApJ...756L...8G,2013ApJ...770..118M,2017xru..conf...87G}; UV, \citealt{1984A&AS...58..705W,1991MNRAS.250..780B,2014A&A...565A..33P}; optical, \citealt{2008ApJ...673..864J,2014A&A...567A.106L,2008PASA...25...69B}; IR, \citealt{1995ApJ...445..716D, 2001ApJ...556..181D,2011ApJ...740...34C}; radio, \citealt{2009ARA&A..47...27K,2014ApJ...783..130R,2014ApJ...794...90K}), different types of objects (\citealt{2009A&A...499..473H,2010MNRAS.402..713S,2011ApJ...733L..43M,2014MNRAS.437.1549B,2014IAUS..298....7C,2016AstL...42....1B,2018IAUS..334..378W}), and many others.

As shown in early \citep{1982PASJ...34..365Y,1983MNRAS.202.1025G} and confirmed by later studies \citep{2001ApJ...553..184C,2001MNRAS.322..426O,2020A&A...637A..96C}, our Galaxy clearly shows a double exponential vertical density profile, the individual components of which can be attributed to the stellar thin and thick discs. In this study, we use this photometric (geometric) definition to separate the thin and thick discs which have significantly different photometric scale heights. The chemical definition of these components is also widely used in the literature and refers to the chemical composition of the stars: $\alpha$-enhanced stars (corresponding to the thick disc component) are more vertically extended versus $\alpha$-poor stars in the thin disc, where $\alpha$ refers to the average abundance of the elements Mg, Si, and Ti \citep{2011A&A...535L..11A,2014A&A...562A..71B,2014A&A...564A.115A,2014ApJ...796...38N}. Also, the stars in the thin and thick discs in the solar neighbourhood are distinguished kinematically \citep{2003A&A...410..527B,2005A&A...433..185B,2006MNRAS.367.1329R}, and by combination of kinematics, metallicities and stellar ages \citep{1998A&A...338..161F,2008MNRAS.388.1175H}. The origin of the thin and thick discs in our Galaxy and in external galaxies \citep{2002AJ....124.1328D,2006AJ....131..226Y, 2017ApJ...847...14E,2018A&A...610A...5C,2019A&A...629A..12M} is still under debate \citep[see e.g.][and Sect.~\ref{sec:discussion}]{2004ApJ...612..894B,2003ApJ...597...21A,2009MNRAS.399.1145S,2009ApJ...707L...1B}.

Interestingly, the Galactic disc scale length $h_\mathrm{R}$ of about 2-2.5~kpc (for both the thin and thick discs according to different estimates, \citealt{2016ARA&A..54..529B}), seems to be too small if we take into account the rather large radius of the galaxy (no strong evidence for a cut-off at $R\gtrsim15-30$~kpc has been found, \citealt{2002A&A...394..883L,2009MNRAS.392..497S,2010ApJ...718..683C,2014A&A...567A.106L, 2020A&A...637A..96C}). This is particularly strange as the majority of the galactic discs are truncated at $3.5-4\,h_\mathrm{R}$ \citep[see e.g.][]{1979A&AS...38...15V,2004MNRAS.355..143K,2007A&A...466..883V}. Here we should note, however, that the disc scale length of the Galaxy has a large uncertainty in the literature \citep[see][and references therein]{2016ApJ...831...71L} -- it spans from 1.8 to 6~kpc! Therefore, the reliability of its disc scale length estimates still raises concerns.

The Galactic bulge and bar also have a complex structure \citep[see e.g.][]{2015MNRAS.450.4050W,2017MNRAS.465.1621P}. Near-infrared observations of the COBE satellite \citep{1992ApJ...397..420B} revealed the boxy nature of the bulge \citep{1995ApJ...445..716D}, which was later confirmed by the 2MASS all-sky map \citep{2006AJ....131.1163S}. In many studies, an X-shape structure in the central part of the Galaxy was identified. The double-peaked distribution of metal-rich red clump giants, which are used as standard candles, points to an X-shape structure in the centre \citep{2010ApJ...721L..28N,2010ApJ...724.1491M,2011AJ....142...76S,2013MNRAS.435.1874W,2015MNRAS.447.1535N}. Also, based on unWISE \citep{2014AJ....147..108L}  3.4$\mu$m and 4.6$\mu$m wide field images, \citet{2016AJ....152...14N} identified and quantified \citep{2017MNRAS.471.3988C} the X/peanut(P)-shaped structure. However, \citet{2016A&A...593A..66L}, \citet{2018IAUS..334..318L} and \citet{2019A&A...627A...3L} do not confirm the discovery of the X-shape structure based on the population of the double red clump.
In addition to that, as stressed by \citet{2017ApJ...836..218L}, oxygen-rich Mira variables (also used as standard candles) match a boxy, but not an X-shape bulge. \citet{2018IAUS..334..263H} do not confirm the result by \citet{2016AJ....152...14N} on the X-shape structure by analysing a residual map for the bulge model using the same unWISE wide-field map. Therefore, there is no consensus on the presence of an X-shape structure in our Galaxy, although it is well-seen in many edge-on galaxies \citep[see e.g.][]{2006MNRAS.370..753B,2014MNRAS.444L..80L,2017MNRAS.471.3261S}.

In this paper, we aim at studying the (thin+thick) disc+bulge structure of the Milky Way and exploit the same integrated unWISE 3.4~$\mu$m photometry as used by \citet{2016AJ....152...14N}, \citet{2017MNRAS.471.3988C} and \citet{2018IAUS..334..263H}. However, compared to those studies, here we take into account the projection effects and the location of the observer inside the Galaxy and consider the whole range of Galactic longitudes $0\degr \leq l < 360\degr$. We accurately fit the 3D structure of the Galaxy by increasing the complexity of the model, including the observed over-densities and disc warping and flaring. In the future paper, this will allow us to directly compare the structural properties of the Milky Way with other spiral galaxies, for which photometric decomposition based on an integrated NIR photometry has been carried out in a similar way. Also, we use this photometry in attempt to constrain the properties of the Milky Way bar (its position angle, major axis, and the properties of the X-shape structure, which we unambiguously detect by matching with our accurate N-body simulations).

This paper is organised as follows. In Sect.~\ref{sec:data}, we describe the image preparation for the Galaxy using the unWISE photometry. In Sect.~\ref{sec:method}, we describe our fitting method. The results of our modelling are presented in Sect.~\ref{sec:results}.
We discuss our results in Sect.~\ref{sec:discussion} and make final conclusions in Sect.~\ref{sec:conclusions}.

\section{The data}
\label{sec:data}
For our modelling of the Milky Way structure, we exploit a specially created mosaic image of the Galaxy with the aid of the same approach as in \citet{2016AJ....152...14N}. \citet{2014AJ....147..108L} reprocessed the data from the Wide-Field Infrared Survey Explorer (WISE,  \citealt{2010AJ....140.1868W}) in the W1-W4 bands (3.4-22~$\mu$m), presented in the AllWISE data release, which were specially convolved for detecting isolated point sources. In contrast, \citet{2014AJ....147..108L} presented a new set of coadds of the WISE images (``unWISE'') that have not been blurred and, thus, retain the intrinsic resolution of the initial data\footnote{\url{http://unwise.me/}}. In a new release of the unWISE data, \citet{2017AJ....153...38M} updated the coadds using the data from the previous release and those obtained after the reactivation of the WISE satellite. Also, in these new coadds, the impact of the scattered light from the Moon and other artefacts was significantly reduced. The coadd images have been sky-subtracted, therefore no additional reduction of these data is required.

For our analysis of the Galaxy structure, we only acquire the unWISE W1-band ($3.4\mu$m) images. In this band, the emission is dominated by the old stellar population which makes up the bulk of the stellar mass of the Galaxy \citep[see e.g.][]{ 2013AJ....145....6J,2014ApJ...788..144M}: this waveband is not sensitive to star-forming regions and young stars, therefore it is best suited for studying the structure of the stellar components. More importantly for our study, dust extinction plays a minor role and does not drastically affect the stellar emission beyond +/- several degrees from the mid-plane (see also below). We do not use the WISE W2 ($4.6\mu$m) band as it is close to the W1 band and does not convey any new information on the stellar structure. Also, it is more affected by the nonstellar emission of hot dust than the WISE W1 band (according to \citealt{2001ApJ...554..778L}, approximately $1/13$ of the observed emission in the WISE W1 band is the dust glow versus $1/8$ at $4.6\,\mu$m, see also \citealt{2017MNRAS.471.3988C}).
On the other hand, the emission in the WISE W1 band contains some emission from the $3.3\mu$m polycyclic aromatic hydrocarbon (PAH) feature \citep{2007ApJ...657..810D,2008ARA&A..46..289T}: the C-H stretching mode near $3\mu$m produces a rich spectrum with the dominant peak at $3.3\mu$m, which is usually accompanied in galaxies by a weaker emission feature at $3.4\mu$m. We neglect this effect in our study as the PAH $3.3\mu$m-feature emission is mostly concentrated within the thin layer $|b|<2.5\degr$ \citep{1994A&A...286..203G,2013PASJ...65..120T}, which is masked out in our study. 
The WISE W3 ($12\mu$m) band also contains significant PAH features and the WISE W4 ($22\mu$m) band is dominated by continuum emission from hot dust grains. In principal, these two WISE W3, W4 bands would help to correct the WISE W1 image for the emission of PAHs and hot dust. However, in 2010 August and September, the W3 and W4 channels, respectively, became unusable due
to the depletion of cryogen. At present, the W3- and W4-band unWISE coadds contain artefacts around large, bright structures. Therefore, the last two bands are not usable in this study as well.

The final stacked image of the Galaxy in the WISE W1 band is created in the following way. We download all unWISE fields which cover the following area in the celestial sphere (in Galactic coordinates): $0\degr\leq l  < 360\degr$ and $-30\degr\leq b \leq +30\degr$. As our primary interest in this paper is studying the structure of the thin and thick discs and the bulge, we do not consider the area with $|b|>30\degr$, which is primarily related to the stellar halo. Initially, we created an empty image with a size of $480\times960$~pixels with a world (Galactic) coordinate system using a Hammer-Aitoff projection which covers the entire sky with an average pixel scale within the Galactic plane as $360\degr/960\mathrm{pix}=0.375\degr/\mathrm{pix}$ (at the image centre). For each pixel in the created image, we find all fields in the unWISE database which overlap with the area covered by that pixel (in Galactic coordinates). We estimate the total flux within that area and assign the measured flux to the pixel value.

Following \citet{2017MNRAS.471.3988C}, we reduce the contamination from bright sources in our image, such as bright nearby stars, supergiants, star clusters, and nearby galaxies (the Andromeda galaxy, the Large Magelanic Cloud etc.). To ``filter out'' these bright sources, we went through all pixels in our image for searching significantly offset pixels with an intensity value $>2\sigma$ above or below the median found for the adjacent pixels. The outliers so determined were then replaced by their symmetric counterparts on the opposite side below or above the disc mid-plane, respectively. By so doing, we reduce the noise level to obtain a more smoothed image of the Galaxy which is important for the subsequent Galaxy fitting. For the same purpose, we additionally smooth the processed image with a Gaussian kernel size of $\sigma=2$~pixels. This is done to obtain a more diffuse light distribution in our Galaxy image. We thoroughly tested that this smoothing did not influence the results of our fitting described in Sect.~\ref{sec:method}.

\begin{figure*}
\label{fig:MW_map}
\centering
\includegraphics[width=15cm]{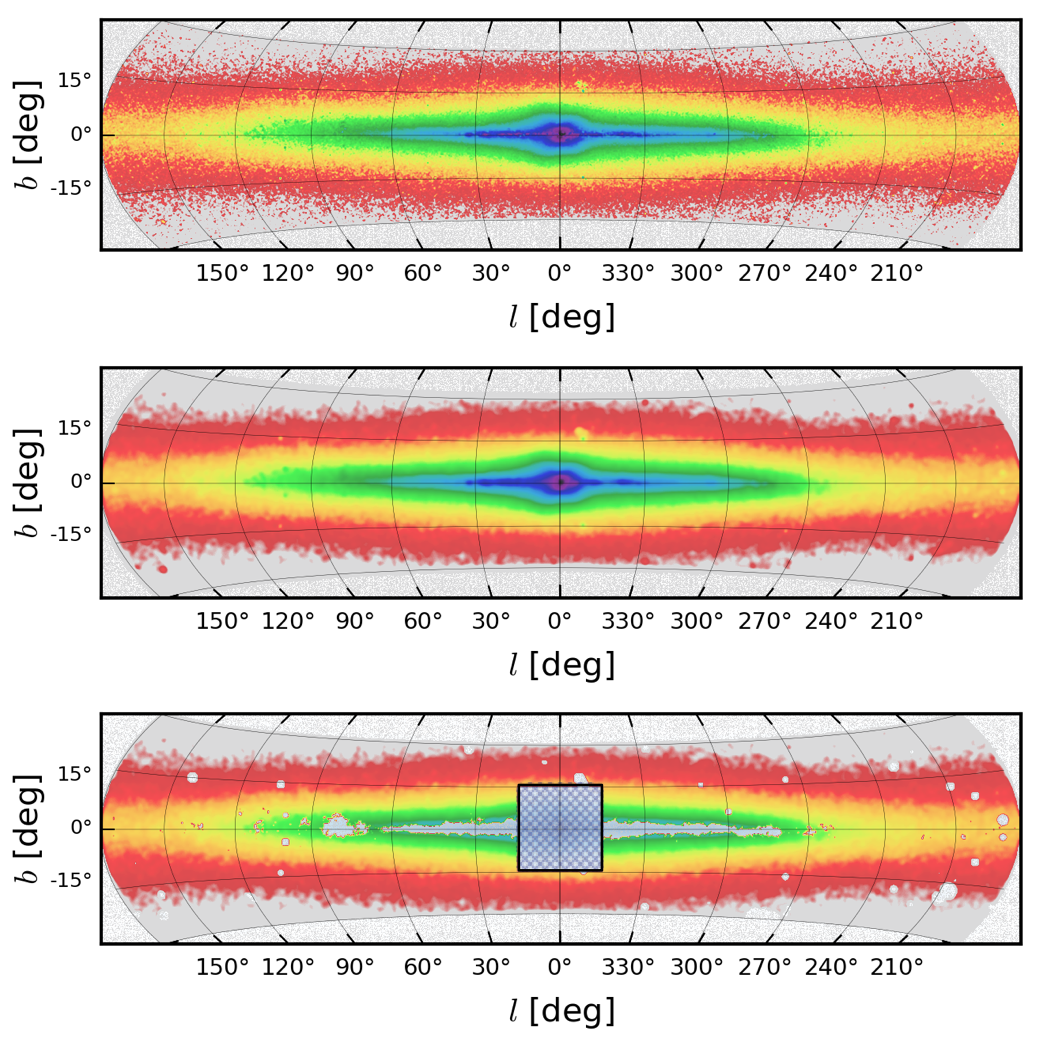}
\caption{Final images of the Galaxy, based on the unWISE data: the image with symmetric replacement of outliers (top), the same but smoothed (middle), and the same but with the mask superimposed (bottom). The masked bulge is shown inside the black shaded box.}
\end{figure*}

Finally, we should take into account the extinction due to the interstellar dust, which is well-visible in the mid-plane of the Galaxy in our image. Unfortunately, the existing extinction and reddening maps \citep[e.g][]{1998ApJ...500..525S,2009MNRAS.395.1640R,2014A&A...571A..11P,2015ApJ...798...88M} are not well-consistent \citep{2017ASPC..510...75G,2017MNRAS.472.3805G,2018MNRAS.475.1121G,2020MNRAS.tmp.2594G,2020MNRAS.tmp.2586G}, especially within the dust layer toward the Galactic plane. Therefore, we decided to mask all pixels in which the corrected intensity significantly differs from the uncorrected one: $|I(x,y)_\mathrm{cor}-I(x,y)|/I(x,y)\geq0.3$. Here, we apply the dust map by \citet{1998ApJ...500..525S} and calculate the extinction in the W1 band as $A_{W1}=0.039\,A_V$ using the extinction curve by \citet{2007ApJ...663..320F} and $R_V=3.1$. In these pixels, which generally occupy $|b|\lesssim2.8\degr$ along the minor Galaxy axis and $|l|\lesssim157.5\degr$ along the major Galaxy axis, the optical thickness is extremely large ($\tau_\mathrm{V}\gg1$) and, thus, the dust extinction is significant even at 3.4~$\mu$m. Our mask (depicted in Fig.~\ref{fig:MW_map} by a light colour) is even more conservative than in other similar studies ($|b|<2-2.5\degr$, \citealt{2013A&A...552A.110G,2016PASA...33...25Z,2018IAUS..334..263H}). In addition to the masking, we could correct the Galaxy image itself for the dust extinction using the already adopted extinction map for creating the mask. Instead, we use this extinction map to correct for dust the emission models of the stellar components which are produced in Sect.~\ref{sec:method} without taking into account any dust component in our modelling.

In addition to the mask of the dust lane, we mask out all contaminants which are still present in the filtered image (see Fig.~\ref{fig:MW_map}). Also, in our main step of the Galaxy fitting (see Sect.~\ref{sec:method}), we manually mask the bulge within a box with a size of $30\degr$. With such a mask, we can only consider disc components, fully neglecting the bulge presence outside the masked region (we specially consider the bulge region in Sect.~\ref{sec:bulge_model}). The final image with the superimposed masks used is shown in Fig.~\ref{fig:MW_map} (bottom plot).

As the fitting procedure requires the computation of the \chisquare{}-value which depends on the pixel noise (see Sect.~\ref{sec:method}), we created a pixel-wise noise map by calculating a standard deviation of the pixel values that are neighbours to the given one. In this case, pixels that are located in regions with a high flux variation have lower weights during the fitting procedure.

\section{The method}
\label{sec:method}

Our Galaxy is a composite structure and it has several recognizable components: the thin and thick discs, the boxy/peanut-shaped (B/PS) bulge, and the stellar halo \citep{2008A&ARv..15..145H,2012ARA&A..50..251I,2013pss5.book..271R,2013A&ARv..21...61R,2020ARA&A..58..205H}. In this study we limit ourselves to modelling the disc and bulge components, whereas a stellar halo component is not included in our models. The fact that we neglect the halo in our modelling should not, by any means, significantly affect our decomposition results as this component is very faint in our Galaxy ($\sim1$\% of the total stellar mass, \citealt{2008ApJ...680..295B}).

The strategy of our modelling of the main Galaxy components is as follows. We start with a very simple model consisting of a single stellar disc, while the bulge is masked. We then construct a model, consisting of thin and thick discs. We complicate the model by adding the flaring and warping effects. Finally, we unmask the bulge and use the final model for the discs and several N-body simulations for a Milky Way-like galaxy, where a similar X-structure occurs, to constrain the size and orientation of the observed bulge in our Galaxy (see Sect.~\ref{sec:bulge_model}). We briefly describe this methodology below.

In order to describe the surface brightness distribution of a three-dimensional (3D) axisymmetric stellar disc, we use a 3D luminosity density model where the radial profile of the luminosity density is exponential \citep{1940BHarO.914....9P,1970ApJ...160..811F} and the vertical profile is either exponential \citep[see e.g.]{1989ApJ...337..163W,1999A&A...344..868X,2011MNRAS.414.2446M,2012MNRAS.426.1328B} or sech$^{2}$ (isothermal-sheet disc, \citealt{1942ApJ....95..329S,1950MNRAS.110..305C,1981A&A....95..105V,1981A&A....95..116V,1982A&A...110...61V}). In a cylindrical coordinate system $(R, z)$ aligned with the double-exponential disc (where the disc mid-plane has $z = 0$), the luminosity density $j(R,z)$ is given by
\begin{equation}
j(R,z) = j_{0} \; \mathrm{e}^{-R/h_\mathrm{R}-|z|/h_\mathrm{z}}\,,
\label{exp_disc}
\end{equation}
where $j_{0}$ is the central luminosity density of the disc, $h_\mathrm{R}$ is the disc scale length and $h_\mathrm{z}$ is the vertical scale height.

In the case of an isothermal disc, its luminosity density is described by the following profile:
\begin{equation}
j(R,z) = j_{0} \; \mathrm{e}^{-R/h_\mathrm{R}} \sech^{2} (|z|/2h_\mathrm{z})\,.
\label{sech2_disc}
\end{equation}

The difference between the two kinds of the disc is that the isophotes of an exponential disc are more discy (``sharper'') and flatter than those of an isothermal one with the same geometric scales. \citet{1989ApJ...337..163W} showed that the vertical profiles in redder bands for IC\,2531 have an excess at small heights $z$ over the isothermal model and thus are better fitted by an exponential law. Recently, \citet{2020MNRAS.494.1751M} considered the outer shape for a sample of edge-on galaxies. They found that the outer shape of the stellar disc in only some galaxies is better described by a more discy profile ($\approx$exp), whereas the outer isophotes for most galaxies look less discy ($\approx$sech$^2$) and can be purely elliptical or even boxy. They concluded that the shape of the outer isophotes of edge-on galaxies can be related to their merger history and its intensity. In this study we consider both kinds of the stellar disc and then conclude which model better (if any) matches the observed 2D profile of our Galaxy.

From many studies, it is well established that both the Galactic thin and thick discs are flared \citep[see e.g.][]{2000astro.ph..7013A,2002A&A...394..883L,2006A&A...451..515M,2016A&A...591L...7B,2017A&A...602A..67A,2018IAUS..334..378W,2019ApJ...871..208L}. This means that the disc scale height increases with galactocentric distance. To account for this effect, we modify the function from \citet{2002A&A...394..883L} to describe the dependence of the scale height $h_\mathrm{z}$ on the Galactocentric radius $R$:
\begin{equation}
h_\mathrm{z}(R) =h_\mathrm{z}(R_\mathrm{flare})\,\mathrm{e}^{(R-R_\mathrm{flare})/h_\mathrm{R,flare}}\,,
\label{flare}
\end{equation}
where $R_\mathrm{flare}$ is usually taken as $R_{\sun}$. The parameter $h_\mathrm{R,flare}$ controls the strength of the flaring. In our study, we use two types of flaring. In the case of the ``continuous'' flaring,  $h_\mathrm{z}(R)$ starts to change from the very Galaxy centre. In the case of the ``outer'' flaring, $h_\mathrm{z}(R)$ does not change in the inner Galaxy region and starts to change beyond some $R_\mathrm{flare}$:

\begin{equation}
h_\mathrm{z}(R) = \begin{cases} h_\mathrm{z}(R_\mathrm{flare}), & \mbox{if } R<R_\mathrm{flare} \\  h_\mathrm{z}(R_\mathrm{flare})\,\mathrm{e}^{(R-R_\mathrm{flare})/h_\mathrm{R,flare}}, & \mbox{if } R\geq R_\mathrm{flare} \end{cases}\,.
\label{outer_flaring}
\end{equation}

As can be seen in Fig.~\ref{fig:MW_map}, the Galactic disc exhibits a slight warp, which was first noted in \citet{1993AJ....105.2127C} for the stellar disc and in \citet{1993A&AS...99..105M} for the {\sc Hi} disc. The location of the mean mid-plane warped disc is, however, mainly limited within $|b|<3-4\degr$ (see e.g. fig.~9 in \citealt{2006A&A...451..515M}). 
Nevertheless, this affect becomes more apparent in the 2nd and 3rd quadrants in the Galactic coordinate system (i.e.  at $90\degr<l<270\degr$). Therefore, following multiple studies, where this effect has been taken into account \citep{2000astro.ph..7013A,2002A&A...394..883L,2006A&A...451..515M,2014A&A...567A.106L,2017A&A...602A..67A,2019ApJ...871..208L,2020NatAs...4..590P}, we include it in our models.

We model the geometric shape of the Galactic warp as a vertical displacement of the Galactic disc in the Galactocentric cylindrical coordinates $(R,\phi,z)$, where $\phi$ is the phase angle increasing in the direction of Galactic rotation:
\begin{equation}
z_\mathrm{w}(R,\phi) =\begin{cases} h_\mathrm{w}(R)\,\sin(\phi-\phi_\mathrm{w}) , & \mbox{if } R>R_\mathrm{w} \\ 0 , & \mbox{if } R\leq R_\mathrm{w} \end{cases}\,,
\label{height_warp}
\end{equation}
where $\phi_\mathrm{w}$ is the phase angle of the warp 
and $h_\mathrm{w}(R)$ is a height function specifying the maximum
amplitude of the warp with respect to Galactocentric radius $R$:
\begin{equation}
h_\mathrm{w}(R) =h_\mathrm{0,w}\,\left(\frac{R-R_\mathrm{w}}{R_\mathrm{w}}\right)^{\alpha_\mathrm{w}}\,,
\label{vert_warp}
\end{equation}
where the parameters $h_\mathrm{0,w}$ and $\alpha_\mathrm{w}$ are the amplitude and the rate at which the warp amplitude increases with Galactocentric radius. $R_\mathrm{w}$ is the radius where the warp starts.

Since using only a 2D photometric map of the Galaxy does not allow one to robustly fit all the parameters of a 3D warp,
we decided to fix the phase angle to $\phi_\mathrm{w}=17.6\degr$ \citep{2019NatAs...3..320C} and
$R_\mathrm{w}=9$~kpc \citep{2002A&A...394..883L,2019ApJ...871..208L} in our complex models to minimize the number of free parameters. Also, we should note that the warp parameters are hard to constrain using a 2D Galaxy image only, therefore one should use the results in Sect.~\ref{sec:results} with caution.

In many edge-on galaxies, the surface brightness profiles clearly show a drop-off at the periphery \citep[see e.g.][]{1979A&AS...38...15V,1981A&A....95..116V,1982A&A...110...61V,1994A&AS..103..475B,2002MNRAS.334..646K,2004MNRAS.355..143K,2012ApJ...759...98C,2017MNRAS.470..427P}. In early works \citep{1992ApJ...400L..25R,2003A&A...409..523R}, there was found to be evidence for a stellar disc cutoff at a radius of $\sim14$~kpc ($R_{\sun}=8.5$~kpc). Star counts in the near-infrared revealed the cutoff of the stellar disc at a Galactocentric distance of $15\pm2$~kpc \citep{1996A&A...313L..21R} ($R_{\sun}=8$~kpc). \citet{2011ApJ...733L..43M} found
an edge to the disc at $R = 13.9\pm0.6$~kpc ($R_{\sun}=8$~kpc). 
However, as shown by \citet{2014IAUS..298....7C}, the findings on the stellar disc truncation in the Milky Way are erroneous either because the dataset used was biased or because the warp and flare was confused with the cut-off.
Moreover, as shown in many more recent studies \citep{2006A&A...451..515M,2007A&A...464..909B,2010MNRAS.402..713S,2014A&A...567A.106L,2020A&A...637A..96C}, at least up to 20-30~kpc there is no strong evidence for a truncation in either
old or young stellar populations. Nonetheless, in this paper we use the cut-off radius $R_\mathrm{tr}=31$~kpc in all models of the stellar discs. Initially, we attempted to fit the models with infinite truncation radii, but the generated discs were too extended and did not follow the observations well at $90\lesssim l \lesssim270$. Therefore, we decided to fix this value at 31~kpc, according to \citet{2018A&A...612L...8L} who spectroscopically confirm the presence of the disc stars out to such distances.

In our complex models, we also fit two strong inhomogeneities in our wide-field image. They may be related to the spiral arms or to the Canis Major and Monoceros ring over-densities \citep[see e.g.][]{2006A&A...451..515M}, which are most visible across the large regions at $200\degr<l<280\degr$ \citep{2006MNRAS.366..865B} and $120\degr<l<240\degr$, $-30\degr<b<+40\degr$ \citep[][]{2016ApJ...825..140M}, respectively.

We make use of the Monte Carlo radiative transfer code {\small SKIRT} \citep{2011ApJS..196...22B,2015A&C.....9...20C,2020A&C....3100381C} to perform radiative transfer calculations in the 3D simulation space. With {\small SKIRT}, a sequence of discrete photon packages, which propagate through the interstellar medium, is generated. Each photon package is emitted by a ``star'', which can be located at some distance within the Galaxy, following the specific density distribution for the given component model.
{\small SKIRT} provides a special class (``instrument'') for fitting a galaxy structure when the observer is located within it and taking into account the projection effects when analysing an all-sky map. This allows us to fully treat our wide-field image of the Galaxy. Although {\small SKIRT allows one to include a dust component to take into account dust scattering and absorption in the interstellar medium, we do not employ such approach since at 3.4$\mu$m the parameters of such a dust component would be poorly constrained due to the significantly less extinction as compared to the UV or the optical (see e.g. \citealt{2016A&A...592A..71M,2018A&A...616A.120M}). Instead, we correct an image of the stellar emission model of our Galaxy, produced by {\small SKIRT}, for the observed Galactic extinction (using one of the extinction maps, as described in Sect~\ref{sec:data}).}

Using {\small SKIRT}, we create a model with the values of the free parameters
obtained at the current step of the minimisation process. Then we
apply the dust reddening map $E(B-V)$ by \citet{1998ApJ...500..525S} to correct the model image for the interstellar extinction: $I_\mathrm{mod}-A_{W1}$, where $A_{W1}=0.039\,A_V=0.039\,R_V \cdot E(B-V)$ and $R_V=3.1$. After that, we calculate
the \chisquare{}-value:
\begin{equation}
  \chisquare \; = \; \frac{1}{N_{DOF}} \sum_{i = 1}^{N} w_{i} \, (I_{\mathrm{obs}, i} \, - \, I_{\mathrm{mod}, i})^{2}\,.
\end{equation}
Here, individual pixels in the weight map have values of $w_{i} = m_{i} / \sigma_{I,i}^{2}$, where $m_{i} = 1$ for the valid
pixels and $m_{i} = 0$ for the masked pixels. $\sigma_{I,i}$ is the measured noise (see Sect.~\ref{sec:data}) at each pixel in the Galaxy image, and
$N_{DOF}$ is the number of degrees of freedom which is defined as the number of unmasked pixels
of the image minus the number of the free parameters of the model.
The obtained \chisquare{}-value is used to asses the goodness of a fit for the current model and compare it with the previous ones. The minimisation algorithm we use in our modelling is the differential evolution search method \citep{Storn1997} which does not require any initial values of the fitting parameters, but only the intervals which limit the optimal parameters.

To estimate the uncertainties of the derived parameters, we launch the decomposition procedure repeatedly several times. Since the differential evolution algorithm starts with a random state, it allows one to run the model trough different evolution paths and see the difference. To add more noise to the process, we also randomly mask 25\% of the image pixels. After one hundred of iterations, we compute a standard deviation of the results for individual runs.

Throughout our paper, we use the following scaling parameters for the location of the Sun within the Milky Way: the
vertical distance from the Galactic mid-plane $z_{\sun}=25$~pc and the distance from the Galactic centre
$R_{\sun}=8.2$~kpc \citep{2016ARA&A..54..529B}.

\section{Modelling the disc components}
\label{sec:results}
We apply the above described method to various models, starting from the simplest one-disc model without any structural features to the most complex model, which accounts for many observational details seen in the wide-field Galaxy image under study.

\begin{figure*}
\label{fig:MW_models_single}
\centering
\includegraphics[width=18cm]{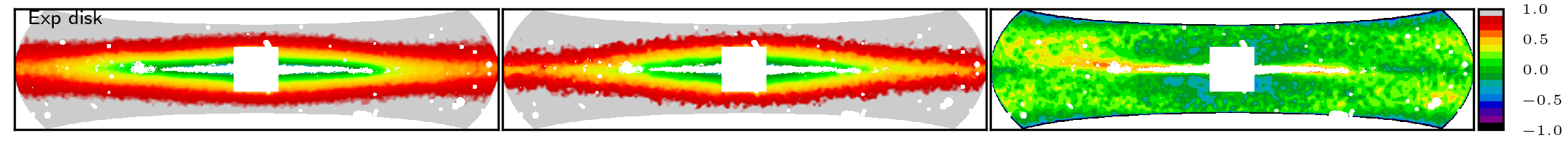}
\includegraphics[width=18cm]{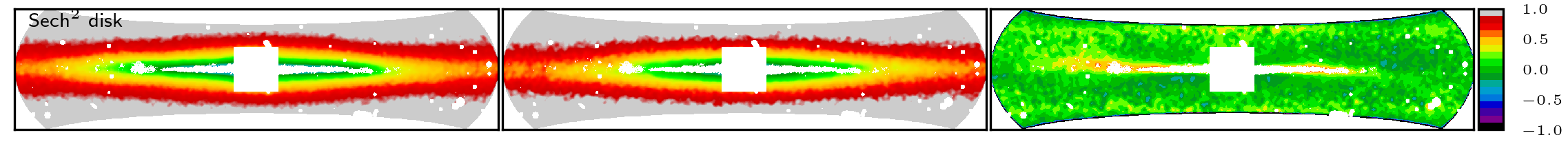}
\includegraphics[width=18cm]{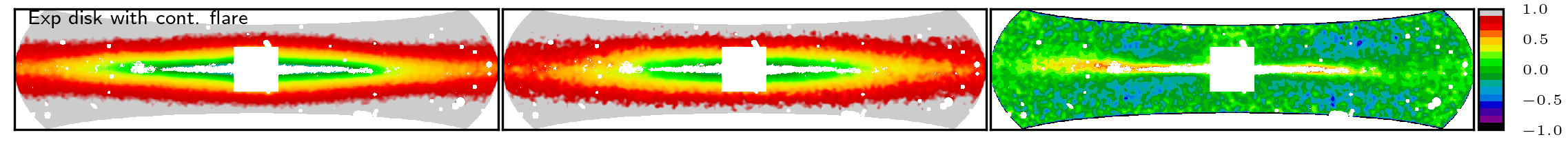}
\includegraphics[width=18cm]{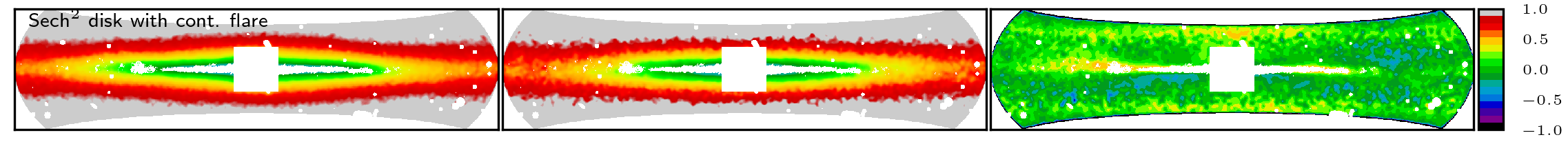}
\includegraphics[width=18cm]{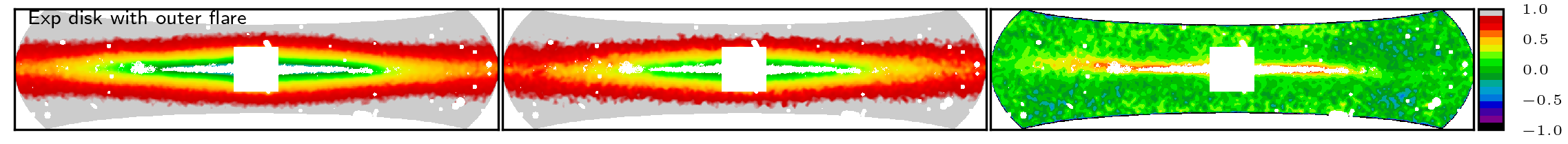}
\includegraphics[width=18cm]{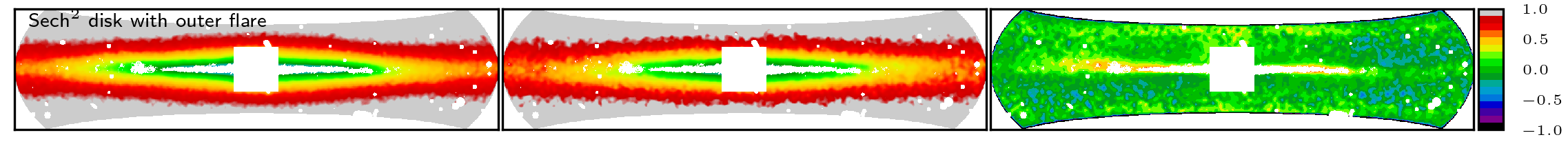}
\caption{Galaxy models with a single disc. The left-hand column always represents the same Galaxy image, the
middle column contains the corresponding fits, and the right-hand panel shows the residual images, which indicate the relative
deviation between the fit and the image.}
\end{figure*}

\begin{figure*}
\label{fig:MW_models_double}
\centering
\includegraphics[width=18cm]{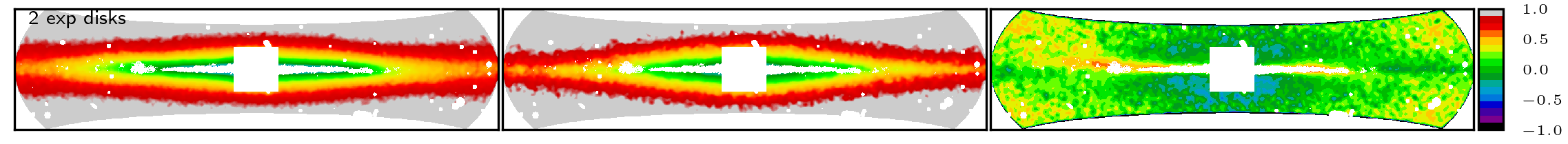}
\includegraphics[width=18cm]{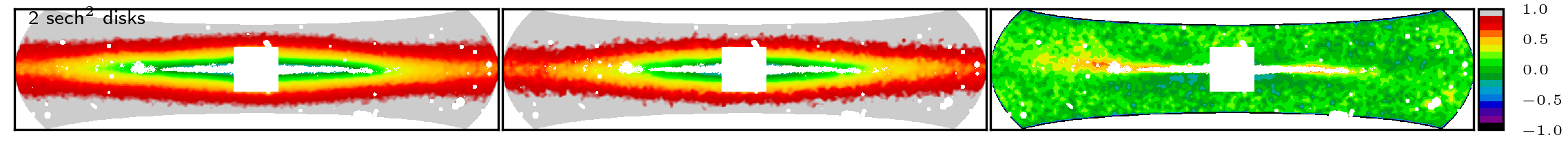}
\includegraphics[width=18cm]{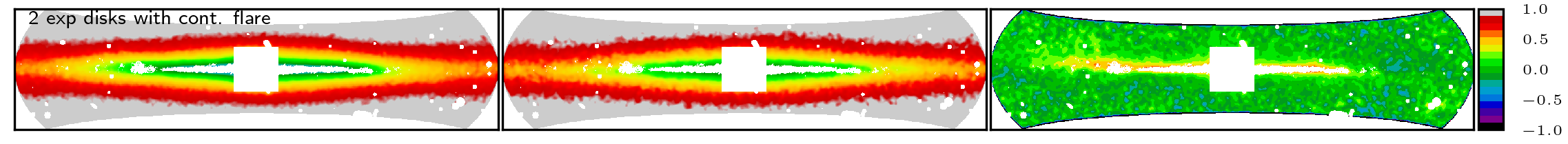}
\includegraphics[width=18cm]{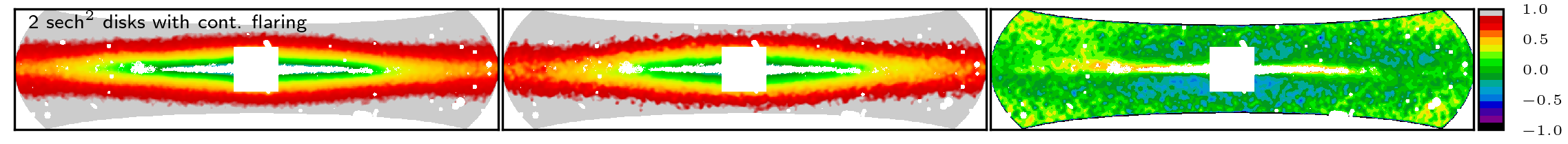}
\includegraphics[width=18cm]{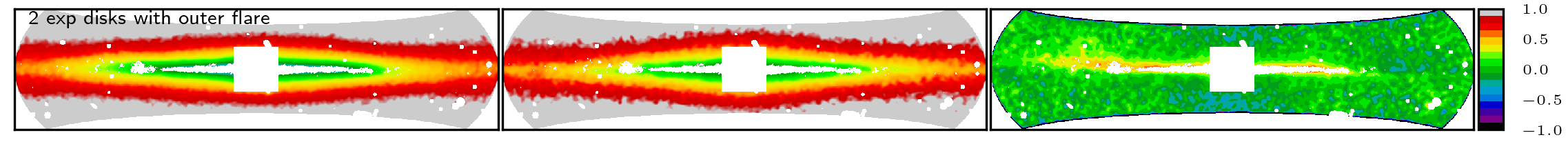}
\includegraphics[width=18cm]{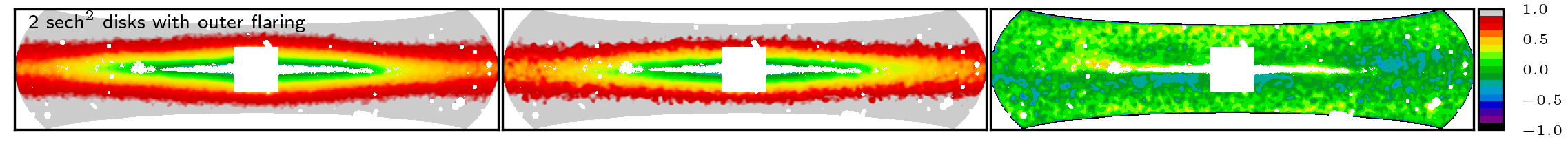}
\includegraphics[width=18cm]{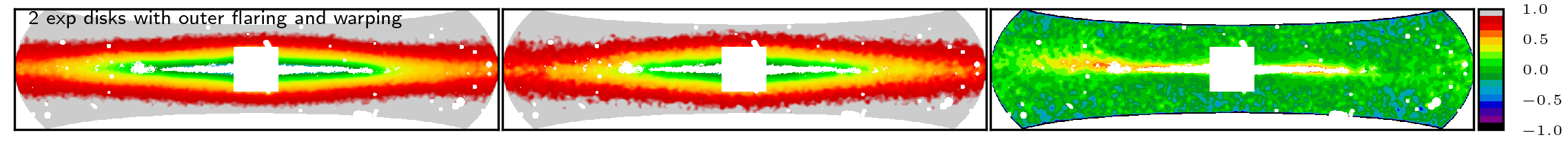}
\includegraphics[width=18cm]{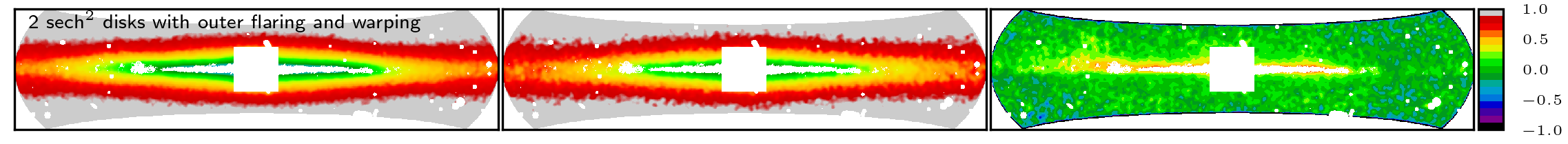}
\includegraphics[width=18cm]{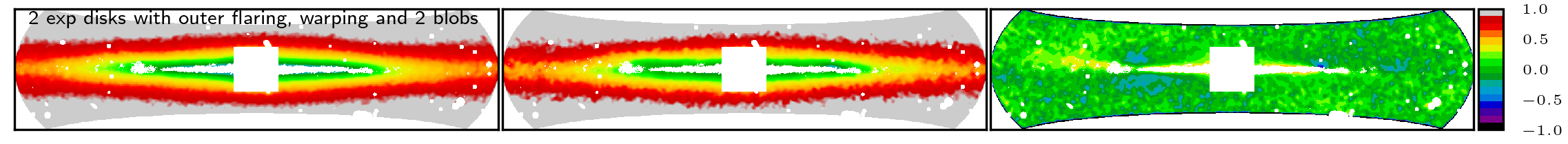}
\includegraphics[width=18cm]{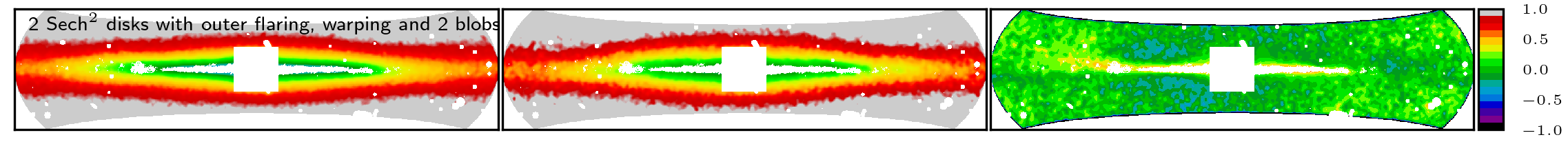}
\caption{The same as in Fig.~\ref{fig:MW_models_single}, but for the two-disc models.}
\end{figure*}

In Table~\ref{tab:fitting_results.tab}, we present the results of our fitting for different models.
In Figs.~\ref{fig:MW_models_single} and \ref{fig:MW_models_double}, one can find the comparison between the Galaxy image and one-disc and two-disc models, respectively. The residual images in these figures (the right-hand columns) show the relative deviation between the fit and the image, that is $(I_\mathrm{obs}-I_\mathrm{mod}) / I_\mathrm{obs}$. Apart from the free fit parameters, we also list the \chisquare{} values which indicate the goodness of the fit. Below we briefly describe each set of models starting from the simplest to the most complex and, presumably, most accurate model. We consider two kinds of the stellar disc: exponential (\ref{exp_disc}) and isothermal (\ref{sech2_disc}).

\begin{table*}
 \centering
 \begin{minipage}{180mm}
  \centering
  \parbox[t]{180mm} {\caption{Results of the fitting for the different models used. By * we denote those parameters which were kept fixed during the fitting in a particular model.}
  \label{tab:fitting_results.tab}}
  \begin{tabular}{cccccccccccccc}
  \hline
  \hline
   $h_\mathrm{R,t}$ & $h_\mathrm{z,t}$ & $R_\mathrm{flare,t}$ &  $h_\mathrm{R,flare,t}$ &  $h_\mathrm{R,T}$ & $h_\mathrm{z,T}$ & $R_\mathrm{flare,T}$ &  $h_\mathrm{R,flare,T}$ & $R_\mathrm{w,t}$ & $\alpha_\mathrm{w,t}$ & $\phi_\mathrm{w,t}$ & $h_\mathrm{0,w,t}$ & $L_\mathrm{T}/L_\mathrm{t}$  & \chisquare{}  \\
    (pc)   &  (pc)    &       (pc)                    &            (pc)                    &   (pc)    &   (pc)   &         (pc)                 &           (pc)                      &
    (pc) &                      &   (deg) &  (pc)    &                  &         \\ \hline
\multicolumn{14}{c}{Single exponential-disc model}\\
3248 &  407   &   --- & --- & --- & --- & --- & --- & --- & --- & ---& --- & --- & 1.26 \tabularnewline
\multicolumn{14}{c}{Single sech$^2$-disc model}\\
3532 &  521  &   --- & --- & --- & --- & --- & --- & --- & --- & ---& --- & --- & 0.93 \tabularnewline
\multicolumn{14}{c}{Single exponential-disc model with continuous flaring}\\
3209  &  579   &   $0^{*}$ & 11226 & --- & --- & --- & --- & --- & ---& --- & --- & --- & 1.32 \tabularnewline
\multicolumn{14}{c}{Single sech$^2$-disc model with continuous flaring}\\
2717 &  451  &   $0^{*}$ & 18852 & --- & --- & --- & --- & --- & ---& --- & --- & --- & 1.12 \tabularnewline
\multicolumn{14}{c}{Single exponential-disc model with outer flaring}\\
3188  &  454   &   $R_{\sun}^{*}$ & 9195 & --- & --- & --- & --- & --- & ---& --- & --- & --- & 1.21 \tabularnewline
\multicolumn{14}{c}{Single sech$^2$-disc model with outer flaring}\\
3023 &  475  &   $R_{\sun}^{*}$ & 15698 & --- & --- & --- & --- & --- & ---& --- & --- & --- & 0.98 \tabularnewline
\hline
\multicolumn{14}{c}{Two exponential-disc model}\\
2988  & 331  & ---  &--- & 4363 & 770 & --- & --- & --- & --- & --- & --- & 0.52 &  1.62 \tabularnewline
\multicolumn{14}{c}{Two sech$^2$-disc model}\\
2602  & 369  & ---  &--- & 4674 & 922 & --- & --- & --- & --- & --- & --- & 0.45 & 0.94 \tabularnewline
\multicolumn{14}{c}{Two exponential-disc model + continuous flaring}\\
3276  & 318 &   $0^{*}$  & 7973 & 2658 & 1081 & $0^{*}$& 11417 & --- & --- & --- & --- & 0.57 & 1.09 \tabularnewline
\multicolumn{14}{c}{Two sech$^2$-disc model + continuous flaring}\\
3281  & 386   & $0^{*}$  &11883 & 2694 & 1049 & $0^{*}$ & 20910 & --- & --- & --- & --- & 0.91 & 1.46 \tabularnewline
\multicolumn{14}{c}{Two exponential-disc model + outer flaring}\\
2931  & 304   & $R_{\sun}^{*}$  & 8286 & 3018 & 961 & $R_{\sun}^{*}$& 17378 & --- & --- & --- & --- & 0.41 & 1.12 \tabularnewline
\multicolumn{14}{c}{Two sech$^2$-disc model + outer flaring}\\
2316  & 271   & $R_{\sun}^{*}$  &10930 & 3133 & 687 & $R_{\sun}^{*}$ & 12860 & --- & --- & --- & --- & 0.81 & 1.16 \tabularnewline
\multicolumn{14}{c}{Two exponential-disc model + outer flaring + warping}\\
2294 & 282   & 8062  & 8368 & 3157 & 659 & 7644 & 9810 & $9000^{*}$ & 0.88 & $17.6^{*}$ & 383 & 0.70 & 1.14 \tabularnewline
\multicolumn{14}{c}{Two sech$^2$-disc model + outer flaring + warping}\\
2558  & 212   & 6323  & 7712 & 2755 & 878 & 10431 & 7883 & $9000^{*}$ & 0.94 & $17.6^{*}$ & 507 & 0.98 & 0.98 \tabularnewline
\multicolumn{14}{c}{Two exponential-disc model + outer flaring + warping+ 2 blobs}\\
2548  & 252   & 7959 & 7064 & 3219 & 708 & 8480 & 11361 & $9000^{*}$  & 1.57 & $17.6^{*}$ & 219 & 0.71 & 0.83\tabularnewline
\multicolumn{14}{c}{Two sech$^2$-disc model + outer flaring + warping+ 2 blobs}\\
2516  & 238   & 8311 & 6760 & 2793 & 979 & 9589 & 9694 & $9000^{*}$ & 0.85& $17.6^{*}$ & 163 & 0.83 & 0.94 \tabularnewline
  \hline\\
  \end{tabular}
  \end{minipage}
 \end{table*}

\begin{table}
 \centering
 \begin{minipage}{80mm}
  \centering
  \parbox[t]{80mm} {\caption{Final results of the modelling. By `Averaged model' we mean the average values of the parameters for all two-disc models listed in Table~\ref{tab:fitting_results.tab}. By * we denote those parameters which were kept fixed during the fitting.}
  \label{tab:final_results.tab}}
  \begin{tabular}{cccc}
  \hline
  \hline
Parameter                     & Units &  Final model        & Averaged model
\\ \hline
 $h_\mathrm{R,t}$                          &   pc   & $2548\pm191$  & $2731\pm365$   \tabularnewline
 $h_\mathrm{z,t}$                          &   pc   & $252\pm35$      & $296\pm56$       \tabularnewline
 $R_\mathrm{flare,t}$     &   pc   & $7959\pm1677$  & $7664\pm906$                       \tabularnewline
 $h_\mathrm{R,flare,t}$&   pc   & $7064\pm994$  & $8622\pm1824$\tabularnewline
 $h_\mathrm{R,T}$                          &   pc   & $3219\pm417$  & $3246\pm704$    \tabularnewline
 $h_\mathrm{z,T}$                          &   pc   & $708\pm142$    & $869\pm154$    \tabularnewline
 $R_\mathrm{flare,T}$     &   pc   & $8480\pm2323$ & $9036\pm1225$                      \tabularnewline
 $h_\mathrm{R,flare,T}$&   pc   & $11361\pm969$ &$12664\pm4364$\tabularnewline
 $R_\mathrm{w,t}$                         &   pc   & $9000^{*}$ & $9000^{*}$                      \tabularnewline
 $\alpha_\mathrm{w,t}$               &         & $1.57\pm0.34$   &  $1.06\pm0.34$                     \tabularnewline
 $\phi_\mathrm{w,t}$                  &   deg  & $17.6^{*}$   & $17.6^{*}$                     \tabularnewline
 $h_\mathrm{0,w,t}$                    &   pc   & $219\pm97$       &$318\pm157$  \tabularnewline
 $L_\mathrm{T}/L_\mathrm{t}$                   &          & $0.71\pm0.45$   &$0.69\pm0.20$   \tabularnewline
  \hline\\
  \end{tabular}
  \end{minipage}
 \end{table}

\begin{table*}
 \centering
 \begin{minipage}{180mm}
  \centering
  \parbox[t]{180mm} {\caption{Covariance matrix for the fitted parameters of our final model consisting of two flared discs with a warp and two blobs.}
  \label{tab:cov}}
\centering
\begin{tabular}{|c|c|c|c|c|c|c|c|c|c|c|c|c|c|}
  \hline
  \hline
&$L_\mathrm{t}$&$h_\mathrm{R,t}$&$h_\mathrm{z,t}$&$L_\mathrm{T}$&$h_\mathrm{R,T}$&$h_\mathrm{z,T}$&$h_\mathrm{R,flare,t}$&$h_\mathrm{R,flare,T}$&$R_\mathrm{flare,t}$&$R_\mathrm{flare,T}$&$\alpha_\mathrm{w,t}$&$h_\mathrm{0,w,t}$&background\\
  \hline
$L_\mathrm{t}$&1.00&-0.12&0.10&-0.78&0.29&0.46&-0.11&-0.08&-0.24&0.16&-0.11&0.16&0.16\\
$h_\mathrm{R,t}$&-0.12&1.00&0.32&-0.17&-0.33&0.03&0.17&0.06&-0.25&0.05&-0.06&0.10&-0.26\\
$h_\mathrm{z,t}$&0.10&0.32&1.00&-0.22&0.13&-0.12&-0.01&-0.03&0.28&0.11&-0.03&0.04&-0.20\\
$L_\mathrm{T}$&-0.78&-0.17&-0.22&1.00&0.10&-0.17&0.13&-0.05&0.21&-0.25&0.08&-0.09&-0.43\\
$h_\mathrm{R,T}$&0.29&-0.33&0.13&0.10&1.00&0.38&-0.04&0.09&0.02&-0.07&-0.02&0.08&-0.17\\
$h_\mathrm{z,T}$&0.46&0.03&-0.12&-0.17&0.38&1.00&-0.03&-0.19&-0.13&0.26&-0.04&0.11&-0.28\\
$h_\mathrm{R,flare,t}$&-0.11&0.17&-0.01&0.13&-0.04&-0.03&1.00&-0.16&0.06&-0.02&-0.02&-0.11&0.12\\
$h_\mathrm{R,flare,T}$&-0.08&0.06&-0.03&-0.05&0.09&-0.19&-0.16&1.00&-0.15&-0.13&-0.03&-0.04&0.05\\
$R_\mathrm{flare,t}$&-0.24&-0.25&0.28&0.21&0.02&-0.13&0.06&-0.15&1.00&-0.28&0.01&-0.30&0.03\\
$R_\mathrm{flare,T}$&0.16&0.05&0.11&-0.25&-0.07&0.26&-0.02&-0.13&-0.28&1.00&-0.00&0.06&0.04\\
$\alpha_\mathrm{w,t}$&-0.11&-0.06&-0.03&0.08&-0.02&-0.04&-0.02&-0.03&0.01&-0.00&1.00&-0.02&0.06\\
$h_\mathrm{0,w,t}$&0.16&0.10&0.04&-0.09&0.08&0.11&-0.11&-0.04&-0.30&0.06&-0.02&1.00&-0.08\\
background&0.16&-0.26&-0.20&-0.43&-0.17&-0.28&0.12&0.05&0.03&0.04&0.06&-0.08&1.00\\
  \hline\\
  \end{tabular}
  \end{minipage}
 \end{table*}

\subsection{Simple single-disc models}
\label{sec:one_disc}
First, we fit a single exponential/sech$^2$ disc model to the Galaxy image. As can be seen in Table~\ref{tab:fitting_results.tab}, the disc scale lengths in both models appear to be similar, whereas the scale height of the sech$^2$ disc is 1.3 times larger than that of the exponential disc. The comparison of the \chisquare{}-values and the residual images suggest that the sech$^2$-disc model better follows the observed profile of the Galaxy, especially in the central Galaxy region. 
Also, the sech$^2$-model shows less discy isophotes in comparison with the diamond-like exponential disc. 
Nonetheless the satisfactory results of this modelling, warping and flaring of the disc are immediately seen in both residues. Also, one can see in the residual images that the left and right sides relative to the galaxy centre are not symmetrical. We consider all these features in our two-disc models below. However, we first fit single-disc models with flaring, to ensure that an additional stellar disc is actually required to describe the observed 2D Galaxy profile. We do not include disc warping in our one-disc models as it is only apparent within a few degrees from the Galaxy midplane, while here we aim to fit the overall disc structure, which does not show a significant warp.

\subsection{Single-disc models with a flare}
\label{sec:one_disc_soph}
For each of the disc kinds from the previous models, we add continuous (where the change of the disc scale height starts at the very Galaxy centre outwards) or outer (where the change of the disc scale height is starting at some radius) flaring, which are described by eqs. (\ref{flare}) and (\ref{outer_flaring}), respectively. The results of the modelling lead us to conclude that the models with the outer flare better agree with the Galaxy image, especially in the 1st and 4th quadrants, whereas the regions above and below the bulge are significantly overestimated by the models with the continuous flare. Interestingly, the main parameters of the disc did not change significantly in the models we have considered so far (within 20\%, on average). For all the single-disc models considered, we compute the following average values:  $\langle h_\mathrm{R} \rangle=3153\pm270$~pc and $\langle h_\mathrm{z} \rangle=481\pm61$~pc.

In general, the sech$^2$ discs yield better results. Also, the flare scales for the exponential discs appear to be lower than those for the sech$^2$ discs which is explained by the fact that exponential discs are more discy (i.e. naturally less ``flared'')  than their sech$^2$ analogues, therefore more ``flaring'' (i.e. a lower flare scale) is required for such models to fit the flared Galaxy disc. 

In principle, the models with the outer flare describe the light distribution in the Galaxy fairly well. However, some underestimation of the flux in the mid-plane within $|b|\lesssim10\degr$ is apparent. Also, one can clearly see a warp of the disc within the same range of Galactic latitudes. Obviously, an additional thin disc component is required to mitigate the observed mismatch of the single-disc models and the Galaxy image.

\subsection{Simple two-disc models}
\label{sec:double_disc}
In this set of models, we fit thin plus thick discs to the Galaxy image, for both the exponential and isothermal cases. The retrieved parameters for both kinds of the disc do not differ significantly, except for the scale height of the thick disc, which is 20\% larger for the sech$^2$ profile. The unsatisfactory negative residue in the 1st and 4th quadrants and a quite large \chisquare{}$=1.62$ for the exponential disc suggest that this simple model is worse than those found in the case of a single disc (Sect.~\ref{sec:one_disc_soph}). This serves as evidence that some additional improvements of the two-disc models are required.

In most models, which we describe below (including the simple two-disc models),  the thick disc has a larger radial scale length than the thin disc. Also, the thick-to-thin disc luminosity ratio $L_\mathrm{T}/L_\mathrm{t}$ appears to be larger than 0.5 and close to 1 for most complicated models, which means that the luminosity of the thick disc is almost as high as that of the thin disc. This is an important result of our study which we discuss in Sect.~\ref{sec:thin_thick_discs}.

\subsection{Two-disc models with a flare}
\label{sec:double_disc_flaring}
In addition to the previous models, we take into account disc flaring (continuous and outer) for both the thin and thick discs. The models with an outer flare slightly better match the observations. Interestingly, the goodness of the fits with an outer flare for the single sech$^2$-disc models is better than in the case of the two-disc models with a flare. We also note that the radial scale of the thick disc is smaller for the models with a flare than for the models without it.

\subsection{Two-disc models with flaring and warping}
\label{sec:double_disc_flaring_warping}
Finally, based on the previous models with an outer flare, we add warping to the thin disc to account for the visible
bending along the Galaxy mid-plane. As noted in Sect.~\ref{sec:method}, the warp radius $R_\mathrm{w,t}$ is fixed to 9000~kpc, together with the phase angle $\phi_\mathrm{w,t}=17.6\degr$. Also, in these and next models we do not fix the radius $R_\mathrm{flare}$ where the outer flaring starts. As can be seen in Fig.~\ref{fig:MW_models_double}, the resultant models better reproduce the light distribution in the Galaxy plane where the thin disc dominates.

\subsection{Final model}
\label{sec:final_model}
To fit the observed over-densities on both sides relative to the Galaxy centre, at $l\sim81\degr$ and $l\sim314\degr$, we add two blobs, each represented by a 3D S\'ersic \citep{1963BAAA....6...41S,1968adga.book.....S} function. The results of our decomposition show that both the exponential- and sech$^2$-disc models reproduce the Galaxy image quite accurately. Not only do the residues demonstrate very little deviation or unfitted features, also the uncertainties on each of the derived parameters (computed using bootstrapping, see Sect.~\ref{sec:method}) are modest, mostly within 15\% of the parameter value, with the exception of the thick-to-thin disc luminosity ratio (see Table~\ref{tab:final_results.tab}). We choose the latter exponential-disc model as our final model as it has the best residue and \chisquare{} among all models obtained. We use it for fitting the central component as described in detail in the next Sect.~\ref{sec:bulge}. The choice of our final complex model is in line with other studies where a detailed structural analysis of the Galaxy was carried out (see Sect.~\ref{sec:lit_comparison}). The robustness of the results for our final model is discussed in Sect.~\ref{sec:models_comparison}.

\begin{figure*}
\label{fig:nox}
\centering
\includegraphics[width=18cm]{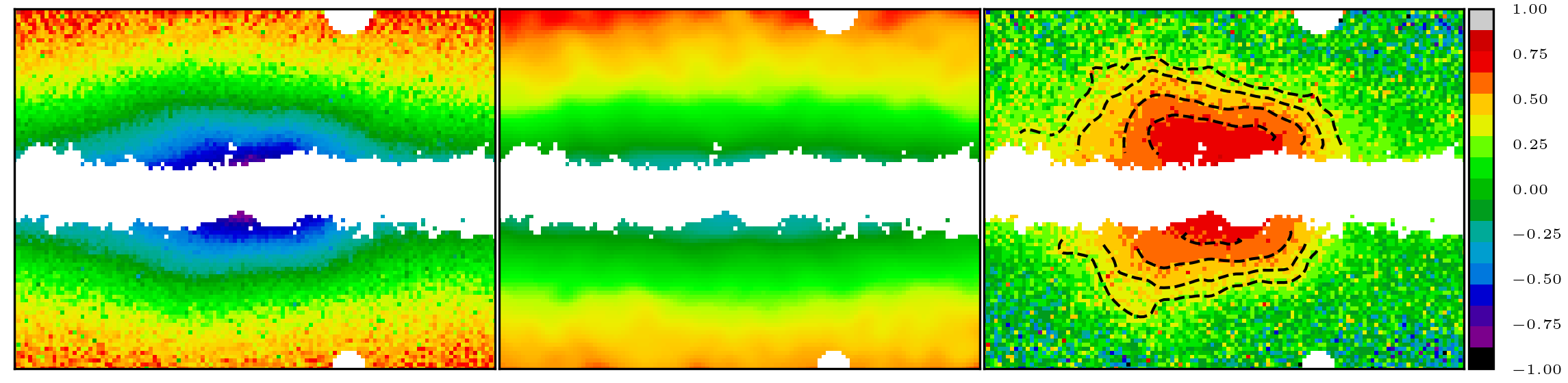}
\caption{The zoomed-in central $30\degr \times 30\degr$ region of the galaxy (left), final model (middle), and relative residual image (right). The white pixels are the mask. The dashed lines represent arbitrarily selected contours which outline the residual X-shape structure.}
\end{figure*}

In Fig.~\ref{fig:nox}, we show the central part of the Galaxy where the bulge dominates. In the residual image (the right-hand plot), one can see an apparent asymmetric structure in the shape of X. This is one of the irresistible conclusions from our modelling. In the next Sect.~\ref{sec:bulge}, we use this final model to constrain the parameters of the X-shape structure in our Galaxy.

In Table~\ref{tab:final_results.tab}, we also list the average values of all derived parameters in our two-disc models. One can see that the main structural parameters of the discs have uncertainties within $\sim20$\% from model to model. We discuss in detail the results of our decomposition in Sect.~\ref{sec:discussion}.

\section{Bulge fitting}
\label{sec:bulge}

\begin{figure*}
\label{fig:bulge_fits}
\centering
\includegraphics[width=18cm]{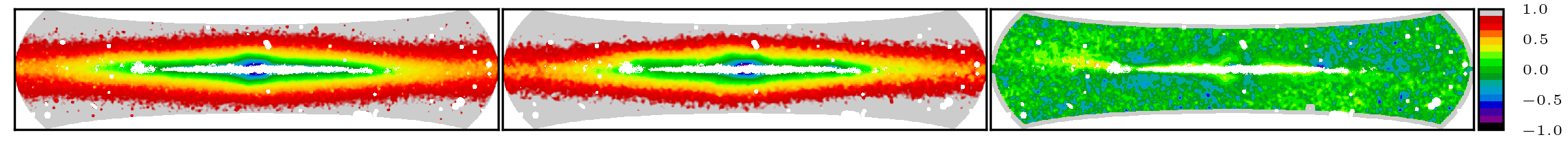}
\includegraphics[width=18cm]{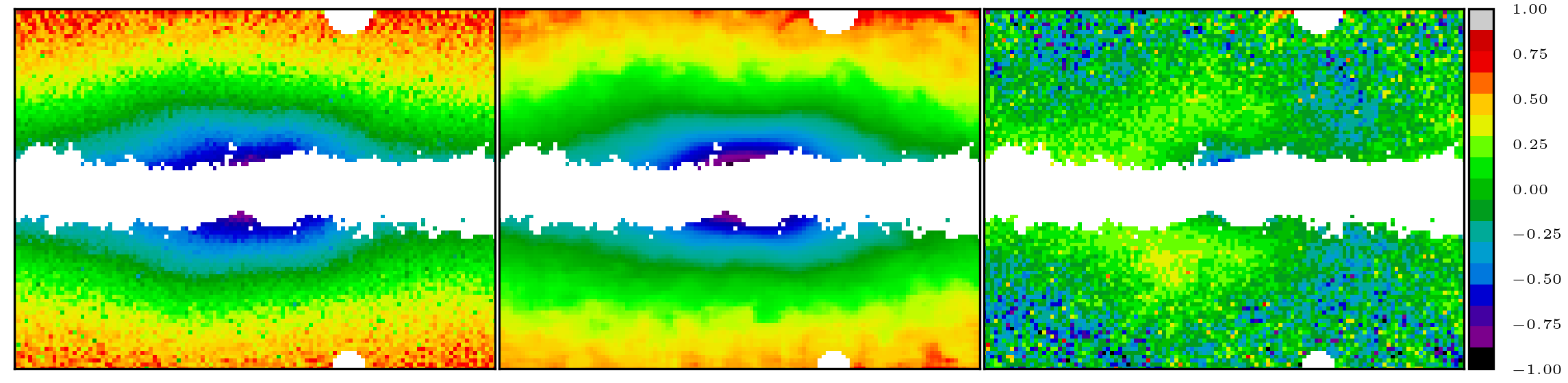}
\includegraphics[width=18cm]{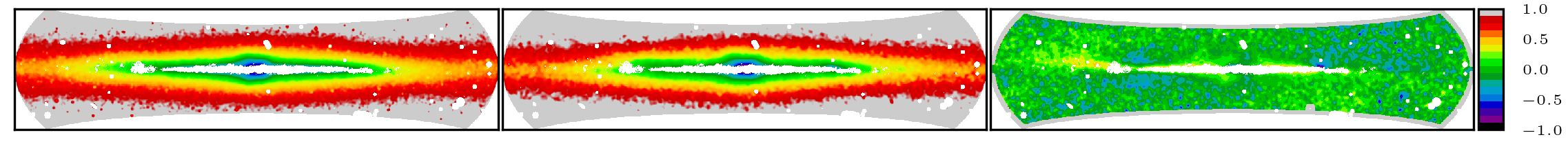}
\includegraphics[width=18cm]{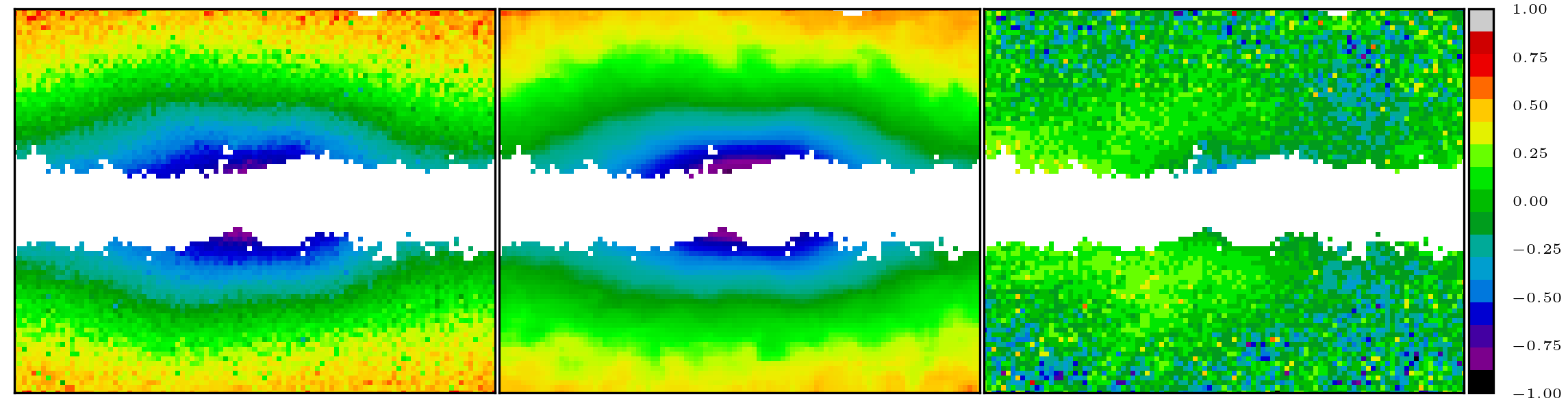}
\caption{Final model with the B/PS bulge component fitted and its zoomed-in central part. \textit{Two top rows}:
		the B/PS bulge model adopted from the initially bulgeless $N$-body model ($M_\mathrm{b}=0$). 
\textit{Two bottom rows}: the B/PS bulge model adopted from the $N$-body model with the initial classical bulge ($M_\mathrm{b}/M_\mathrm{d}=0.2$).}
\end{figure*}

In the previous section, we extracted the disc(s) parameters without fitting the bulge component of our Galaxy. This component was completely masked to avoid the possible influence of the rather complex bulge structure on the disc(s) parameters. Now we turn our attention specifically to the bulge component and describe a procedure which will allow us to estimate some of its parameters. An important point here is that the bulge of the Milky Way is believed to be a vertically extended part of the bar (B/PS bulge, see e.g.\citealt{1990A&A...233...82C,1991A&A...252...75P,2006ApJ...637..214M,2010ApJ...720L..72S}). One of the main problems with studying the Galactic bar is that its major axis orientation is close to the line-of-sight (LoS) orientation \citep[see e.g.][]{1995ApJ...445..716D} and, thus, it is quite hard to distinguish the peanut structure in our Galaxy which is clearly observed in some other edge-on galaxies \citep[e.g.][]{2000A&AS..145..405L,2006MNRAS.370..753B}. As we sit in the disc plane ourselves, the lobes of the bulge are slightly asymmetric with respect to the vertical axis, simply due to the fact that one side of the bar is closer to us. This effect is well distinguished in the asymmetry of the intensity map shown in Fig.~\ref{fig:nox}: on the left side of the bulge, the residue shows a slightly more vertically extended profile compared to the right side.
\par
In principle, we can describe the asymmetric B/PS bulge of the Milky Way by some analytical photometric model in a similar way as it was done for other galaxies \citep[see e.g.][]{2017MNRAS.471.3261S,2020MNRAS.499..462S,2016MNRAS.459.1276C}. We genuinely considered some of these models but found that they perform rather poorly due to the small rotation angle of the bar major axis to the LoS and the overall complexity of the central component. Therefore, we decided to take a different approach and make use of some $N$-body galaxy models from our previous work \citep{2018MNRAS.481.4058S} where a bar and a B/PS bulge naturally arise. We consider two different models with quite different B/PS bulges (the details are given below). In these models, we distinguish particles - ``stars'' that are associated with their B/PS bulges. Next, we use the 3D density distributions of such particles as the basis for the photometric model of the B/PS bulge. We make the 2D intensity maps from such 3D density distributions for different bar viewing angles and different distances from the observer to the centre and try to fit the real observed B/PS bulge component by such images. The advantage of such an approach is that it allows us to account for the B/PS bulge asymmetry which naturally arises due to the fact that the observer is closer to one of the sides of the bar.

\par
\begin{figure*}%
\begin{minipage}[t]{0.95 \textwidth}
\includegraphics[width=1.0\textwidth]{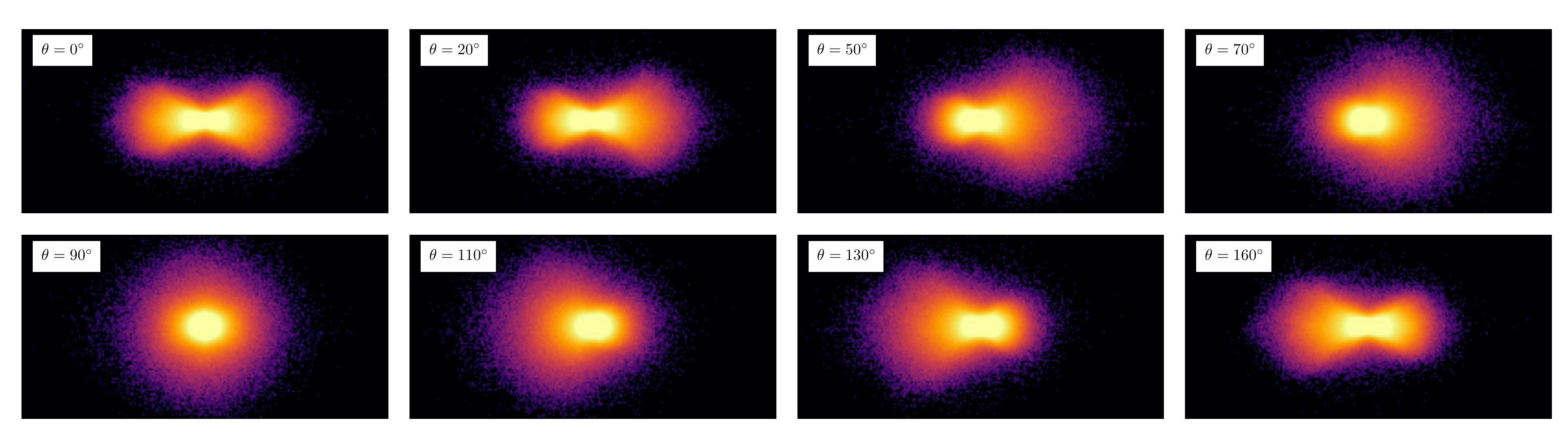}

\includegraphics[width=1.0\textwidth]{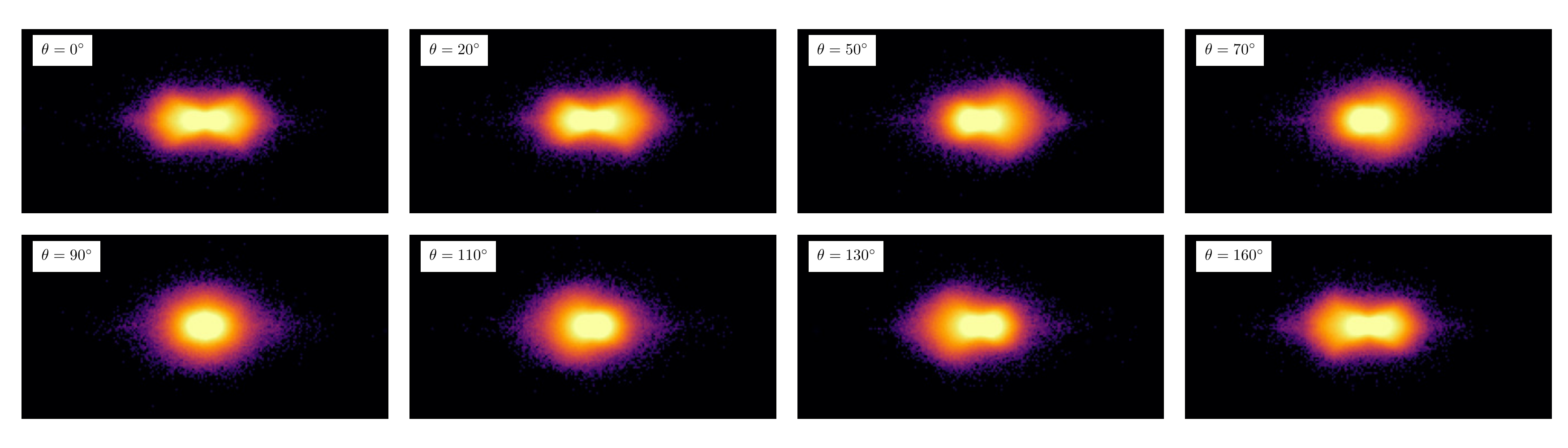}%
\end{minipage}%
\caption{$N$-body bar model viewed from different angles. The top two rows are
for the initially bulgeless model, the two bottom ones -- for the model with an initial classical bulge.}
\label{fig:bar}
\end{figure*}

The details of the described procedure are as follows:
\begin{enumerate}
    \item First of all, the $N$-body models we consider should be characterised. Initially, each of the models has one exponential disc consisting of isothermal sheets in the vertical direction and a spherical dark halo with the profile close to the Navarro-Frenk-White type \citep{1996ApJ...462..563N}. Here we consider two similar models which differ only in that one of the models also initially contains a classical bulge with $M_\mathrm{b}/M_\mathrm{d}=0.2$, where $M_\mathrm{b}$ and $M_\mathrm{d}$ are the total masses of the bulge and the disc, respectively. All other initial parameters of the models, such as the disc scale length, mass, and others are the same. A thorough description for the construction of such models can be found in \citet{2018MNRAS.481.4058S}, so we refer the interested reader to it. Here we should mention that the discs in both models were initially thin $h_\mathrm{z}/h_\mathrm{R}=0.025$. Also, the discs were relatively cool with the Toomre parameter value $Q=1.2$ at $R=2\,h_\mathrm{R}$. The disc and halo components in each of the models were allowed to evolve under the influence of their mutual gravitation. The time evolution of the models was calculated via \texttt{gyrfalcON} code \citep{2002JCoPh.179...27D} which is part of the publicity available \texttt{NEMO} software package \citep{1995ASPC...77..398T}. At about 1~Gyr, the bar forms in the models and, after that, it almost immediately starts to thicken forming a B/PS bulge. After about 4~Gyr, both models evolve only slowly. We chose the specific time moment $t\approx6.5$~Gyr and below we consider the model B/PS bulges for this time moment.
    \item  We distinguish particles - ``stars'' that are associated with the B/PS bulge in each of the models under consideration using typical frequencies of the particles -``stars'', $f_x$, $f_z$ and $f_\mathrm{R}$, where $f_x$ and $f_z$ are the frequencies of oscillations along the $x$- and $z$-axes, respectively, and $f_\mathrm{R}$ is the frequency of cylindrical distance oscillations. This procedure had been already carried out in \citet{2020MNRAS.499..462S} for these models, so here we used their data.
    \item After we found the 3D density distribution of the B/PS bulge in each of the models, we prepared two sets of images (one for each of the considered models) where the B/PS bulge is depicted for different bar viewing angles and distances from the observer to its centre (see Fig.~\ref{fig:bar}). Such images then can be used as the basis for a photometric function of the B/PS bulge. As can be seen from Fig.~\ref{fig:bar}, the initially bulgeless model gives rise to the pronounced peanut-shaped structure with a clear dip of intensity between the B/PS bulge lobes. The model with the classical bulge, on the opposite, has a more compact B/PS bulge with a less pronounced dip of intensity. We note that, although our second model contains the classical bulge, we do not include this component in the B/PS bulge density distribution that we further use to fit the B/PS bulge of our Galaxy. Although from the physical point of view it might be reasonable, we note that the central concentration, which leads to a more compact B/PS bulge, can be built using various sources, not only by a classical bulge but, for example, by a gaseous component too. We utilise the model with the initial classical bulge as the general example of the model with a quite compact B/PS bulge. Also, as our approach is rather new, we are mainly interested in the fact whether or not it can be used to differentiate between the B/PS bulges arising in different physical models of our Galaxy.
    \item We feed the whole set of images to the {\small IMFIT} package \citep{2015ApJ...799..226E}, along with the fixed final model obtained in Sect.~\ref{sec:results}. To do that, we modified the {\small IMFIT} code in such a way, that it accepts a FITS file with an image as a separate component of the fitting model (instead of an analytical function, as it is usually done). The free parameters of such an image-component are coordinates of the centre (i.e. where this component will be placed on the total model image), a scale factor (to control the relative size of the component on the total image), and its total flux (governs the brightness of this component in the resulting model image). The search for the optimal values of this image-component, along with the other classical parameters of the model, is performed in a standard way using a gradient descent technique.
    \par
    \end{enumerate}
    \par
\begin{figure}
    \centering
    \includegraphics[width=1.0\columnwidth]{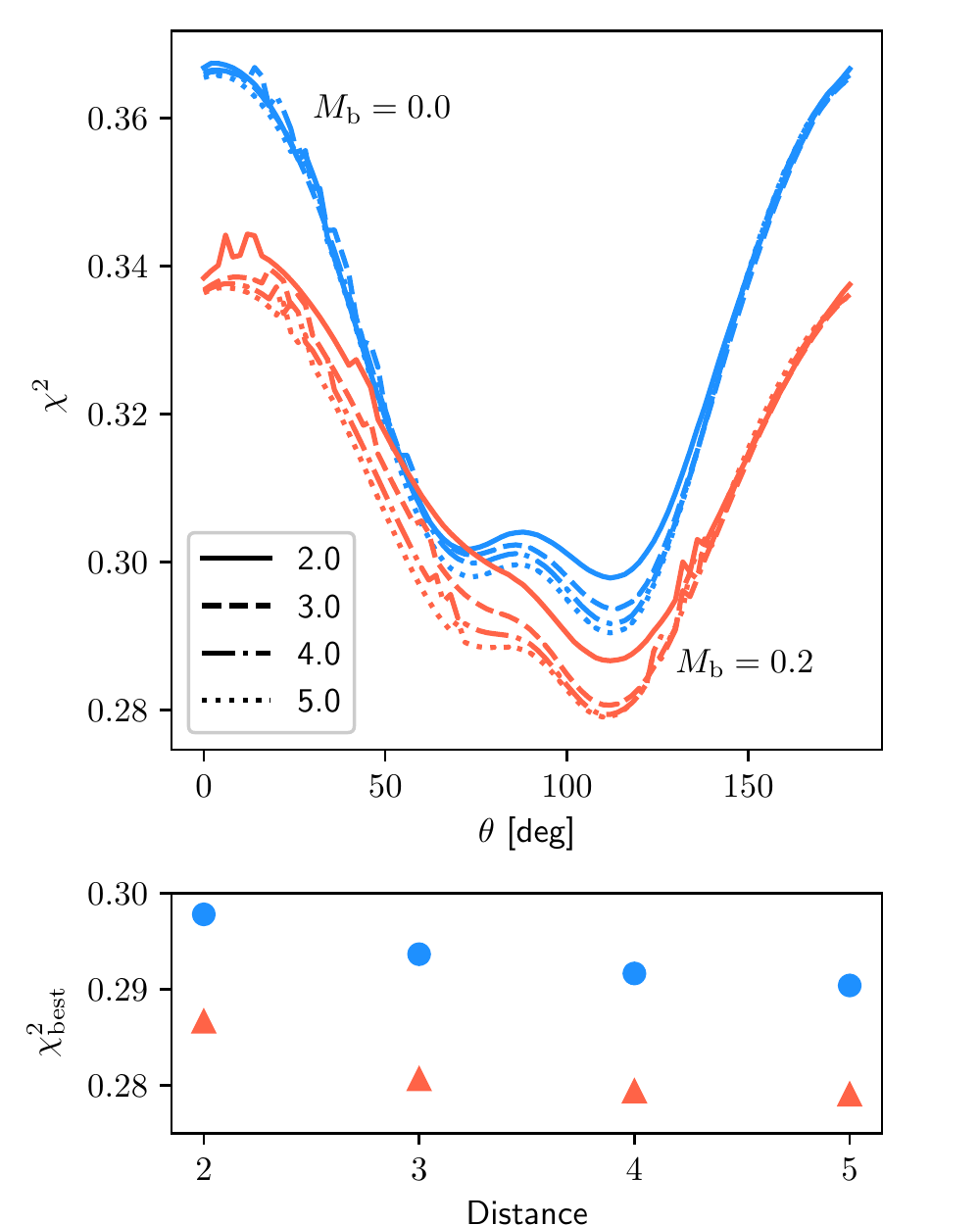}
    \caption{Upper plot: \chisquare{}-value as a function of bar viewing angle for the bulgeless model ($M_\mathrm{b}=0$) and the model with $M_\mathrm{b}=0.2$ for a set of distances from 2.0 to 5.0 in units of the initial disc scale length of the models. Bottom: the best \chisquare{}-values for various bar distances for $M_\mathrm{b}=0$ (closed circles) and $M_\mathrm{b}=0.2$ (open triangles).}
    \label{fig:bar_chi_squared}
\end{figure}

 Fig.~\ref{fig:bulge_fits} shows the best-fit photometric models (for $M_\mathrm{b}=0$ and $M_\mathrm{b}=0.2$) of the Milky Way. In addition to the final model from Sect.~\ref{sec:results}, these models include the B/PS bulge component. As can be seen, our approach indeed allows us to reproduce an asymmetric intensity distribution which we observe in reality. Fig.~\ref{fig:bar_chi_squared} shows how the \chisquare{}-value changes with the bar viewing angle and the distance from the observer to the centre. As can be seen, there is a clear dependence of the \chisquare{}-value on the bar viewing angle with the optimal value $\Theta= 113\degr \pm 2\degr$ for both of the models under consideration. In the present work, we measure the viewing angle in such a way that the zero value corresponds to the bar observed side-on and the angle grows clock-wise when viewed from the north galactic pole. Thus, the optimal value translates into the usual $\phi_\mathrm{bar}=23\degr \pm 2\degr$ if we calculate it from the LoS. Concerning the dependence of the \chisquare{} on distance, we could not find an optimal value of the distance as it becomes degenerate for large distance values. We discuss the obtained results in Sect.~\ref{sec:bulge_model}.

    \par

\section{Discussion}
\label{sec:discussion}

In this section we discuss the robustness of the obtained models in Sect.~\ref{sec:results}, compare the results of our final model with the literature, and comment on the structural parameters, which we derived in this study.

\subsection{Validation of the obtained models}
\label{sec:models_comparison}

The results of our fitting for a variety of models in Sect.~\ref{sec:results} suggest that a single-disc model, even with the inclusion of a disc flare, cannot accurately describe the observed Galaxy structure. Nevertheless, using a single-disc model with flaring yields better results than a simple two-disc model without any additional structural features. We have shown that exploiting a two-disc model with additional structural features improves the residues and the \chisquare{}-values. The principal question on using either an exponential or sech$^2$ function to fit the vertical luminosity density distribution remains open, as neither of the disc types shows a significantly better residual than the other, albeit our final exponential model seem to better match the wide-field image of the Galaxy. Also, the retrieved parameters of the thin and thick discs in these models are well-consistent: $\langle h_\mathrm{R,t} \rangle = 2807\pm387$~pc, $\langle h_\mathrm{z,t} \rangle = 297\pm31$~pc, $\langle h_\mathrm{R,T} \rangle = 3283\pm642$~pc, $\langle h_\mathrm{z,T} \rangle = 836\pm179$~pc for the exponential discs versus $\langle h_\mathrm{R,t} \rangle = 2654\pm367$~pc, $\langle h_\mathrm{z,t} \rangle = 295\pm78$~pc, $\langle h_\mathrm{R,T} \rangle = 3210\pm836$~pc, $\langle h_\mathrm{z,T} \rangle = 903\pm137$~pc for the sech$^2$ discs. 

We also thoroughly investigated the impact of the particular mask choice on the fitting results. To do so, we generated a set of mask images with different properties (with the size of the masked central box region up to $40\degr$; using various dust extinction maps). We carried out additional fitting with the different mask images and found no significant change in the decomposition results.

On the whole, our complex best-fit model with a large number of free parameters can suffer a potential degeneracy. 
To address this problem, we estimated the covariance matrix for the fitted parameters. The normalised matrix is presented in Table~\ref{tab:cov}. As can be seen from the table, the strongest observable correlation is between the thin and thick disc fluxes (the negative anti-correlation with the Pearson coefficient $\rho=-0.78$).
All other correlation coefficients are smaller than 0.5 and, thus, we assume that there is no strong correlation between the other pairs of the parameters. The anti-correlation between the disc fluxes is predictable as if we decrease the luminosity for one disc, the luminosity for  another must increase, so that the total luminosity remains the same. In addition to that, the `average' model in Table~\ref{tab:final_results.tab} demonstrates that the variations of values for the output fit parameters between the different two-disc models are fairly modest. Therefore, we conclude that the parameter degeneracies should not be significant in our modelling.

In Fig.~\ref{fig:bar_chi_squared}, we show the histograms of pixels from the relative residual maps created for different models from Sect.~\ref{sec:results}. As one can see, our final model has a narrow and symmetric Gaussian distribution centred at 0. A similar but wider distribution (with a larger number of pixels with greater uncertainties) is shown for the single-disc model with a flare. In contrast, the other models have asymmetric shapes and their peaks are shifted towards positive values. This indicates that some details in the Galaxy image remained unfitted. Therefore, this once again proves the validity of the final model. The comparison with the literature, where absolutely different data and fitting techniques are used, confirms this conclusion (see below). 

\begin{figure}
    \centering
    \includegraphics[width=1.0\columnwidth]{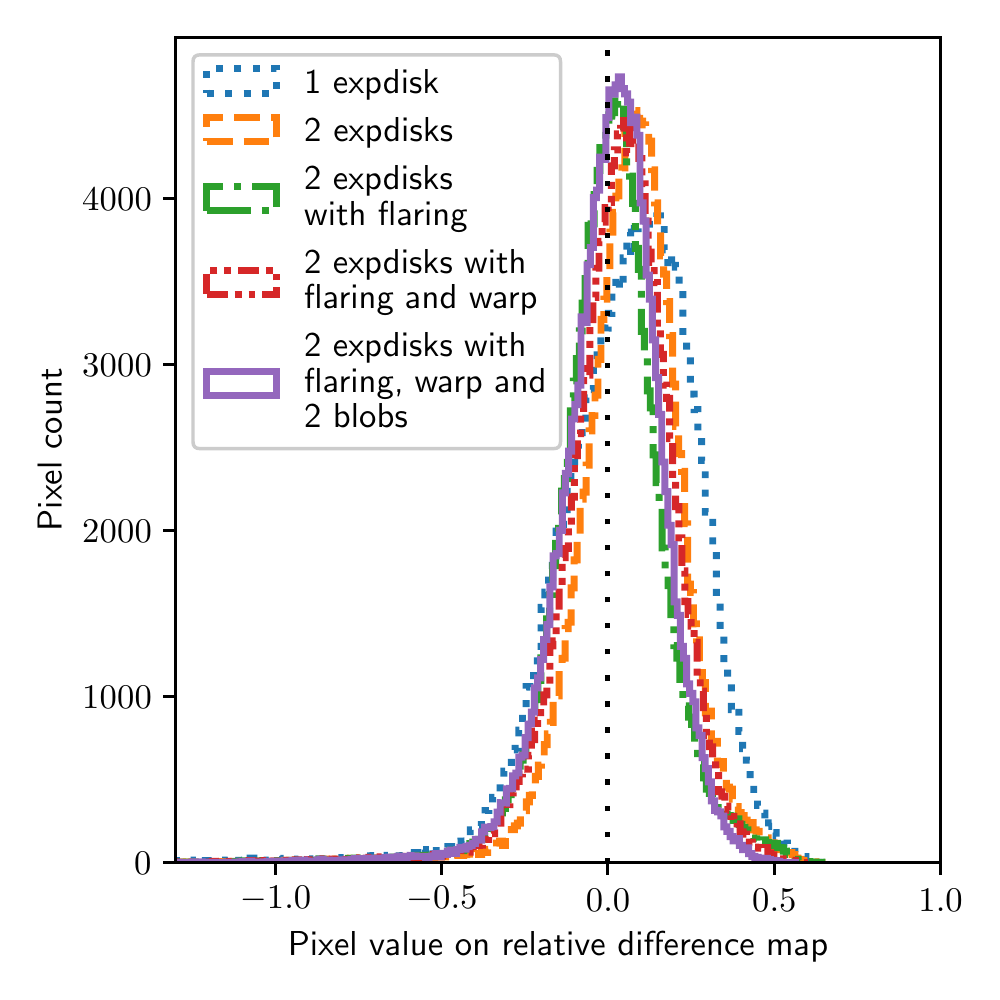}
    \caption{Distributions of the pixel values in the relative residual images for five models: the single-exponential disc model (blue dotted line), the simple model with two exponential discs (orange dashed line), the model with two exponential discs with flaring (green dash-dotted line), the model with two exponential discs with flaring and a warp (red dash-dotted line), and the final model (purple solid line).}
    \label{fig:pix_dist}
\end{figure}

\subsection{Comparison with the literature}
\label{sec:lit_comparison}
In our simple single-disc models, the disc scale length $\langle h_\mathrm{R} \rangle=3.15\pm0.27$~kpc appeared to be slightly larger than computed in \citet{2016ApJ...831...71L} who combined 29 different photometric measurements of the disc scale length from the literature and applied a Bayesian meta-analysis to obtain the improved, aggregate estimate $h_\mathrm{R}=2.64\pm0.13$~kpc.
\citet{2017MNRAS.471.3988C} used a similar unWISE composite image of the Milky Way as we created in this study, but for $-45\degr<l<45\degr$. They found the average scale length of the disc in the eastward and westward directions to be 2.54~kpc.
The estimate of the disc scale height in the case of a single-disc model is less representative and merely shows some general scale height of the Galaxy, given the significant difference between the thin and thick disc scale heights.

Let us now consider our final model in comparison with the results from the literature.
Here, we should point out that the radial scale lengths and the vertical scale heights of the Galactic thin and thick discs still remain poorly constrained (see e.g. table~5 and figure~8 in \citealt{2018MNRAS.479..211M} for the estimated scales of the thick disc in the literature). Therefore, a rather wide range of the values can be determined based on different literature sources: $h_\mathrm{R,t}\sim1.8-4$~kpc and $h_\mathrm{z,t}\sim0.2-0.45$~kpc for the thin disc and $h_\mathrm{R,T}\sim2-4.5$~kpc, $h_\mathrm{z,T}\sim0.5-1.2$~kpc for the thick disc. The possible reasons for these discrepancies have been listed in the Introduction. \citet{2016ARA&A..54..529B} attempted to homogenise the values of the structural parameters for the thin and thick discs:
$h_\mathrm{R,t}=2.6\pm0.5$~kpc,
$h_\mathrm{z,t}=0.3\pm0.05$~kpc, and
$h_\mathrm{R,T}=2.0\pm0.2$~kpc,
$h_\mathrm{z,T}=0.9\pm0.2$~kpc. As can be seen from that review, the thick-disc scale length is shorter than that for the thin disc, though there is a number of studies which claim the opposite result \citep[see e.g.][and below in this section]{2008ApJ...673..864J,2014A&A...567A.106L}. In our modelling, the radial scale length is larger for the thick disc and 1.6 times larger than, on average, found in the literature (see discussion in Sect.~\ref{sec:thin_thick_discs}).

Our scale height of the thick disc $h_\mathrm{z,T}=0.71\pm0.14$~kpc is qualitatively consistent within the marginal uncertainties with $h_\mathrm{z,T}=0.8\pm0.3$~kpc for the already mentioned studies and other studies where the thick disc was investigated: \citet{2012ApJ...755..115B},
\citet{2016ApJ...823...30B},
\citet{2014A&A...569A..13R},
\citet{2014A&A...564A.102C},
\citet{2014A&A...564A.102C},
\citet{2001A&A...373..886R},
\citet{1996A&A...305..125R}, and
\citet{2010ApJ...712..692C}.

The scale length and scale height of the thin disc ($h_\mathrm{R,t}=2.55\pm0.19$~kpc, $h_\mathrm{z,t}=0.25\pm0.04$~kpc), which we derived in this study, perfectly agree with the literature (\citealt{2001MNRAS.322..426O,2008ApJ...673..864J,2014A&A...567A.106L} and more references in \citet{2016ARA&A..54..529B} who analysed 130 refereed papers on disc parameters --  see above).

\begin{figure}
\label{fig:flaring}
\centering
\includegraphics[width=8cm]{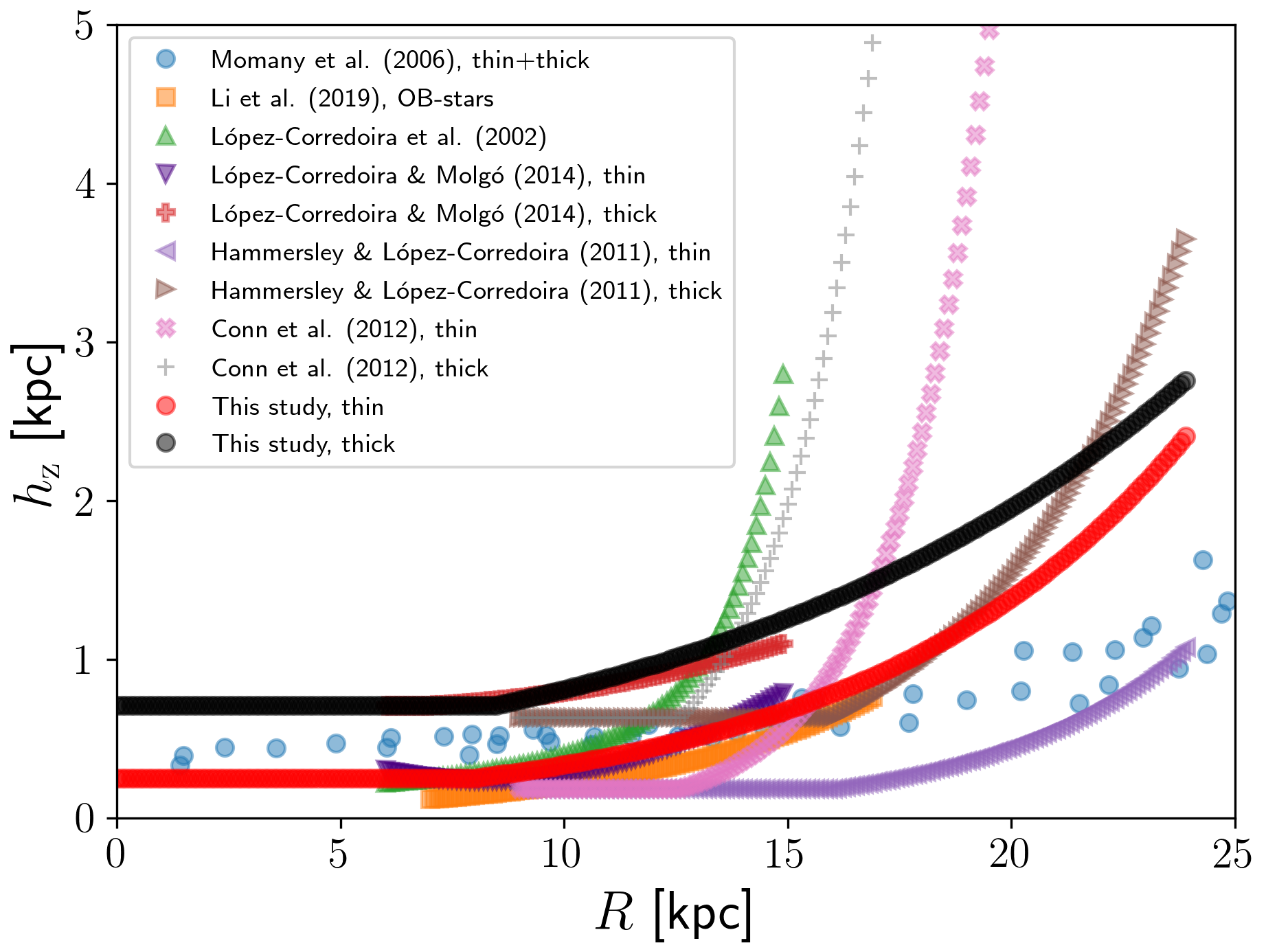}
\caption{Comparison of the disc flare, found in this study, with the literature: \citet{2002A&A...394..883L}, \citet{2006A&A...451..515M}, \citet{2011A&A...527A...6H}, \citet{2012ApJ...754..101C}, \citet{2014A&A...567A.106L}, \citet{2014A&A...569A..13R}, and \citet{2019ApJ...871..208L}.}
\end{figure}

According to our modelling results, the Galaxy exhibits significant outer flaring of the thin and thick discs, starting at approximately the solar radius. Based on the Two Micron All Sky Survey (2MASS, \citealt{1997ASSL..210...25S,2006AJ....131.1163S}) data, \citet{2002A&A...394..883L} found that the flare of the disc starts well inside the solar circle. In our models, we tried to use a continuous flare of the discs, that is there is a decrease of the disc scale heights towards the inner Galaxy, the effect which was pointed out by \citet{2002A&A...394..883L}. However, our modelling with continuous flaring convincingly showed that such models are less consistent with the wide-field Galaxy image as compared to the models with the outer flaring. \citet{2014A&A...567A.106L} used photometric data from the SDSS Sloan Extension for Galactic Understanding and Exploration (SEGUE, \citealt{2009AJ....137.4377Y}) for over 1400~deg$^2$ in off-plane low Galactic latitude regions and derived a flaring in both the thin and thick discs. They identified that the flare becomes quite prominent only at large radii $R>10$~kpc (see their fig.~5 and 6). For OB-stars from the Gaia DR2 \citep{2018A&A...616A...1G}, \citet{2019ApJ...871..208L} found a flare of the outer disc with $h_\mathrm{R,flare}=5.39$~kpc. They concluded that the scale height of OB stars increases more slowly than for red giants.
The thick disc is primarily made up of old stars ($\gtrsim10$Gyr) and many studies showcase that it does not demonstrate a significant flare out to at least 11~kpc \citep{2018MNRAS.479..211M}, 13~kpc \citep{2012ApJ...754..101C}, 15~kpc \citep{2006A&A...451..515M}, or even 16~kpc \citep{2011A&A...527A...6H}. Within that radius, the flare of the thick disc is rather mild. This is generally consistent with what we find in this paper. In Fig.~\ref{fig:flaring}, we show the dependence of the disc scale height(s) on Galactocentric radius for different studies. As one can see, the qualitative consistence between them is poor, which can be explained by the different data used. Nevertheless, the main conclusions from essentially all the studies are that the disc scale height starts to grow beyond the solar radius and that the flaring of the thick disc becomes apparent at large radii.

\begin{figure}
\label{fig:warping}
\centering
\includegraphics[width=8cm]{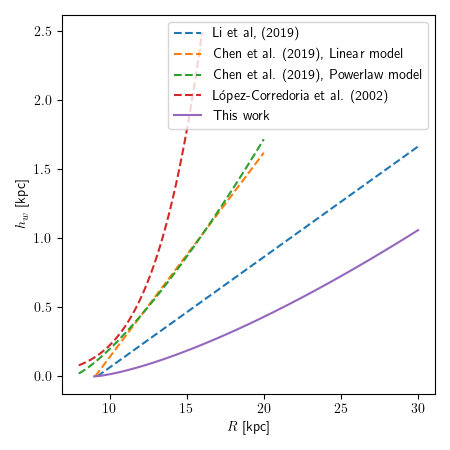}
\caption{Comparison for the disc warp, modelled in this study, with the literature: \citet{2019ApJ...871..208L},  \citet{2002A&A...394..883L},
  \citet{2019NatAs...3..320C}.}
\end{figure}

Fig.~\ref{fig:warping} shows a comparison for the warp of the Galactic disc found in this paper with previous studies (namely, \citealt{2019ApJ...871..208L},  \citealt{2002A&A...394..883L}, \citealt{2019NatAs...3..320C}). As in the case of the flare, the results for the warp display a significant scatter between the studies. The warp model, obtained in this study, does not demonstrate such a rapid growth of $h_\mathrm{w}$ starting from $R_\mathrm{w}$ as found in some previous studies, but has a rather mild transition from the inner, unbend part of the disc to its outskirts. We should note that the parameters of the disc warping in our modelling, according to Table~\ref{tab:fitting_results.tab}, are poorly constrained. This seems quite natural as we only consider a 2D map of our Galaxy, whereas for creating a reliable 3D model of the warp one needs to have a 3D distribution of stars in the Galaxy.

In Sect.~\ref{sec:bulge}, we found the viewing angle of the bar to be $\phi_\mathrm{bar}=23\degr \pm 2\degr$ which is in common with other authors: $20\degr$ \citep{2010ApJ...720L..72S,2012MNRAS.427.1429W}, $22\degr$ \citep{2005MNRAS.358.1309B}, $20-25\degr$ \citep{2014AstL...40...86B,2007MNRAS.378.1064R}, $27\degr$ \citep{2013MNRAS.435.1874W}, $29\degr$ \citep{2013MNRAS.434..595C}.
In contrast, \citet{2017MNRAS.471.3988C}, who used the same unWISE 3.4$\mu$m data, measured the viewing angle $37\degr ^{+7\degr}_{-10\degr}$. In their study, however, they took into account that ``the long bar and the elongated X/P structure of the Milky Way are not distinct and misaligned components, but are different regions of the same structure''.

\subsection{The thin and thick discs in our models}
\label{sec:thin_thick_discs}

In our study, we have only considered the thick and thin discs in the Milky Way from a photometric point of view.
We have demonstrated in Sect.~\ref{sec:results} that our single-disc models with flaring are not able to accurately describe the entire Galaxy structure. This fact goes to prove that, according to the photometric decomposition only, the Galactic structure should be described by two distinct discs. In this case, the scenario, in which the Galactic thick disc naturally arises due to orbital diffusion (stellar radial migration) and/or flaring of the old disc stars without splitting into two distinct stellar components \citep{2008ApJ...675L..65R,2009MNRAS.396..203S,2011ApJ...737....8L,2015ApJ...804L...9M,2017MNRAS.471.2642F,2018A&A...616A..86H}, seems less plausible as compared to some other proposed mechanisms for the thick disc formation (see below). Nevertheless, the thick disc in our final model demonstrates appreciable flaring at large radii. This fact, on the contrary, supports the mechanism that at least a significant fraction of the old stars might form the thick disc through radial migration and flaring \citep[see also][]{2014A&A...569A..13R}.

Apart from its higher elevation, the thick disc is also distinct from its thin counterpart through its chemistry (relatively metal [Fe/H]-poorer and alpha-enhanced) and older age (see e.g. \citealt{2014A&A...562A..71B,2015MNRAS.453..758H,2015MNRAS.453.1855M}). The chemically distinct thin and thick discs are also distinct in their geometrical scales. For example, spectroscopic observations \citep{2012ApJ...752...51C} show that the population of stars with high [$\alpha$/Fe] ratios (which are thus old) has a short scale length as compared to the low-$\alpha$ population (associated with the younger thin disc): $h_\mathrm{R,T}=1.8$~kpc versus $h_\mathrm{R,t}=3.4$~kpc.

Cosmological hydrodynamic simulations convincingly showcase that discs in galaxies grow from the inside out. This has been confirmed observationally for our Galaxy \citep[see e.g][]{2019ApJ...884...99F} and external massive disc galaxies, in which old stellar populations have a smaller radial extent than younger substructures \citep{2004ApJS..152..175M,2007ApJ...658.1006M,2011ApJ...731...10M,2008ApJ...681..244B,2011MNRAS.412.1081W,2015MNRAS.451.2324P}. Therefore, one would expect to derive a longer scale length for a younger stellar population (thin disc) than for the older thick disc \citep[see e.g.][]{2010MNRAS.408.1313L,2013MNRAS.436..625S,2014MNRAS.442.2474M,2015MNRAS.447.2603L,2015ApJ...804L...9M}. According to the other mechanism, thick discs grow through minor mergers \citep{2003ApJ...597...21A,2002ApJ...574L..39G,2006ApJ...639L..13W}. In this case, the
scale length of the thick disc appears to be larger than that of the thin disc, with a prominent flare of the thick disc \citep{2011A&A...530A..10Q,2014A&A...567A.106L}. According to our final model, we find a larger thick disc scale length and a significant flare at the periphery, which seem to be consistent with the latter scenario.

Based on the 3D density of stars of type F8V-G5V from the SEGUE, \citet{2014A&A...567A.106L} also derived a shorter scale length for the thin disc (2.0~kpc) and a larger thick-disc scale length (2.5~kpc). In addition, \citet{2008ApJ...673..864J} found that the number density distribution of M dwarfs in the solar neighbourhood ($D < 2$~kpc) is well fit by two exponential thin and thick discs with scale heights and lengths:
$h_\mathrm{R,t}=2.6$~kpc,
$h_\mathrm{z,t}=0.3$~kpc,
and
$h_\mathrm{R,T}=3.6$~kpc,
$h_\mathrm{z,T}=0.9$~kpc. Their results agree well with the current study.

\citet{2014ApJ...781L..31S} studied the star formation history of the Galaxy using the signatures left on chemical abundances for a sample of long-lived F, G and K stars. They concluded that the formation of the Galactic thick disc occurred sometime between 9.0 and 12.5 Gyr ago, which is coincident with the maximum star formation rate in the Universe \citep{2014ARA&A..52..415M}. Their results imply that the mass of the $\alpha$-enhanced thick disc is comparable to the thin disc \citep[see also][]{2015A&A...578A..87S}. Some earlier studies also demonstrate that the thick disc in the Galaxy is as massive as the thin disc \citep{2008MNRAS.384..173F,2011MNRAS.414.2893F,2012MNRAS.420.1423F}. Recently, \citet{2017A&A...598A..66P} considered two new axisymmetric models for the Galactic mass distribution with a massive and centrally concentrated thick disc and found that they satisfy a number of observational constraints, such as, stellar densities at the solar vicinity, rotation curves etc. For our final decomposition model, the estimated thick-to-thin disc luminosity ratio is $L_\mathrm{T}/L_\mathrm{t}=0.71$. If we adopt the local thick-to-thin disc surface density ratio $f_{\Sigma}=0.38$ from \citet{2014ApJ...781L..31S} and correct it for the retrieved scale lengths of the thick and thin discs, their mass ratio will be $M_\mathrm{T}/M_\mathrm{t}=1.34\pm0.24$. If we use a more conservative estimate $f_{\Sigma}=0.12\pm0.04$ from \citet{2016ARA&A..54..529B}, $M_\mathrm{T}/M_\mathrm{t}=0.42\pm0.16$. Therefore, we can also conclude that the contribution of the thick disc stellar population to the total stellar mass is if not dominant, it is at least very substantial. Our results contrast with what is commonly believed that the mass of the thick disc is only 10-20\% of the mass of the thin disc (\citealt{2011MNRAS.414.2446M,2015MNRAS.448.2934S,2016A&A...593A.108B,2017MNRAS.465...76M}).

The formation of the thick disc can be well explained (at least, for some properties) by the scenario where the Milky Way was gravitationally collapsing from well-mixed gas-rich giant clumps that were sustained by high turbulence, which prevented a thin disc from forming for a time \citep{1998Natur.392..253N,2004ApJ...612..894B,2013A&A...560A.109H,2015A&A...579A...5H,2014ApJ...789L..30L}. This scenario also explains the existence of the old, massive thick disc.

Unfortunately, a fully consistent picture for the thin and thick discs in our Galaxy is yet to be presented, as contemporary data on this matter are not enough to constrain the parameters of these major stellar components and/or our current understanding of the dominant processes in galaxies is far from complete. We can only surmise that different mechanisms for the thick disc formation can be at play at the same time or different periods of time during the Galaxy evolution. For example, the in-situ formation of the thick disc stars at the very beginning of the Galaxy birth and stars formed in the thin disc
and then thickened with time by heating by minor mergers \citep[see e.g.][]{1993ApJ...403...74Q,2008ApJ...688..254K,2008MNRAS.391.1806V,2011A&A...525L...3D} and different thin disc instabilities and structures \citep[see e.g.][]{1984ApJ...282...61S,2010ApJ...721.1878S,2016MNRAS.462.1697A} can be approximately 50-50\% in cosmological high-resolution zoom-in
simulations \citep[][]{2020arXiv200912373P}.

\subsection{The P/X-shape bulge in our model}
\label{sec:bulge_model}

In Fig.~\ref{fig:nox} and then in Sect.~\ref{sec:bulge}, we convincingly showed that the residual structure in the central Galaxy region represents an X-shape structure due to the presence of a bar at a skew angle. Thus, we confirm the results by \citet{2016AJ....152...14N} and \citet{2017MNRAS.471.3988C}, who previously studied a similarly created unWISE Galaxy map and came to the same conclusion.
\par
We found that, by fitting the $N$-body B/PS bulge profile, one can reproduce the real observed asymmetry of the Milky Way B/PS bulge. We considered two different $N$-body models with slightly different B/PS bulges. From Fig.~\ref{fig:bar_chi_squared}, one can see that the models are different in terms of \chisquare{}-values. Thus, with a larger set of models, such an approach can be used to differentiate between various models of the central parts of our Galaxy. We also note that even after fitting the asymmetric B/PS profile, the residue still shows an asymmetric structure (probably, four asymmetric lobes). Comparing the residues of the two investigated models, one can see that this asymmetric structure in the residue becomes less pronounced if we consider a more compact B/PS bulge with a smaller dip of intensity between the X-structure rays. The fact that we still observe a faint asymmetric structure for our second model probably indicates that the B/PS bulge of our Galaxy has an even more compact structure, i. e. the smaller distance between the rays of the X-structure, that we considered in our $N$-body models. At the same time, we should note that we specially estimated the intensity of the obtained residue and found that it contains only about 5\% of the modelled B/PS bulge luminosity. This generally indicates that our model is quite good at the description of the B/PS bulge of our Galaxy.

\par
As to the contribution of the central spheroidal component to the total stellar mass, \citet{2010ApJ...720L..72S} found it to be 10\% at most. \citet{2015ApJ...806...96L} estimated it as $M_\mathrm{B}/M_\mathrm{tot}=0.150$, where $M_\mathrm{B}$ is the mass contained in both the bulge and bar components. In our study, we estimated the B/PS-bulge-to-disc luminosity ratio as 0.14 for the initially bulgeless model and the same 0.14 for the model with the initial central concentration and a more compact B/PS bulge. If we also account for the pixels that form the asymmetric structure observed in the residue in Fig.~\ref{fig:bulge_fits}, the upper bound value will be slightly greater, but still about 0.14, i.e. the influence of that structure is negligible. It is not straightforward to compare the values obtained in the present work with the previous estimates as the methods used are quite different. Nevertheless, our results generally confirm that there is a small room for the bulge of another type to reside in our Galaxy. At least, the residue from Fig.~\ref{fig:bulge_fits} does not show any other pronounced component except the asymmetric structure that is probably due to the B/PS bulge again. As the residue contains only about 5\% of the total B/PS bulge luminosity, this leaves us with B/D$\lesssim0.01$ for other types of the bulge(s).
\par
We should also note that we do no distinguish a plane long bar component, which is discussed in some previous studies~\citep{2016ARA&A..54..529B}. We also do not observe strong density enhancements in the residue picture (Fig.~\ref{fig:bulge_fits}) which can be associated with it. This is probably due to the fact that we masked out the central plane strip contaminated by dust where such enhancements should be observed.

\section{Conclusions}
\label{sec:conclusions}

A study of the Galaxy structure is important for our understanding of how it came into existence.
In this paper, we have attempted to derive the structural parameters of the main Milky Way components using the integrated unWISE 3.4$\mu$m photometry that aid comparison with other galaxies. Our approach contrasts with previous studies based on the integrated unWISE data as follows. First, we utilised the 2D map of the Galaxy in the entire range of Galactic longitudes $0\degr\leq l <360\degr$ and the wide range of Galactic latitudes $|b|\leq30\degr$, where the disc and bulge reside. Second, we considered a set of 3D models of our Galaxy, taking into account that the observer is located within the Galaxy at the solar radius and that the projection of the Galaxy on the celestial sphere should be treated. In our modelling, we consistently considered different models, starting with the simplest model of one disc and ending with a model of two (thin and thick) discs, which include warping and flaring effects. To fit the central component, we used realistic N-body simulations of a Milky Way-like galaxy that self-consistently develops a bar. In the process of evolution, the bar buckles and thickens in the vertical direction and produces an X-shape structure at a skew angle resembling the one observed in our Galaxy. Using the results of these simulations and matching them with the real B/PS bulge in our Galaxy, we were able to estimate the viewing angle of the bar. The values of the parameters derived in this study are summarised in Tables~\ref{tab:fitting_results.tab} and~\ref{tab:final_results.tab}. To sum up, we list the following findings of our study:
\begin{enumerate}
\item Single-disc photometric models of the Galaxy, notwithstanding the inclusion of a disc flaring, do not accurately describe the Galaxy image near the plane and, thus, two-disc models should be considered instead.
\item We examined two kinds of discs in our initial models: exponential and isothermal. We conclude that the final exponential two-disc model with additional structural features slightly better describes the Galaxy structure than does the model with the sech$^2$-discs. In general, the difference between the obtained models with the two kinds of the disc is not significant.
\item On the whole, we find that the disc and bulge parameters in our final model are within the range of the previous measurements in the literature: $h_\mathrm{R,t}=2.55\pm0.19$~kpc, $h_\mathrm{z,t}=0.25\pm0.04$~kpc, $h_\mathrm{R,T}=3.22\pm0.42$~kpc, $h_\mathrm{z,T}=0.71\pm0.14$~kpc, $\phi_\mathrm{bar}=23\pm2\degr$, B/PS-bulge-to-disc luminosity ratio 0.14. These values can be used to constrain the possible models for the formation scenario of the thin and thick discs in the Milky Way.
\item We claim that the mass of the thick stellar disc in our purely photometric study is comparable to the mass of the thin disc ($M_\mathrm{T}/M_\mathrm{t}\sim0.4-1.3$). This result is in line with recent observational and numerical studies \citep[see e.g.][and references therein]{2017A&A...598A..66P}. In the context of hierarchical clustering, our Galaxy presents a major challenge to the standard $\Lambda$CDM paradigm. The existence of the massive thick disc in our Galaxy can be interpreted to favour an internal origin rather than a satellite accretion scenario, although several mechanisms can be at play at the same time. 
\end{enumerate}

In our future study, we are about to compare the photometric parameters of the Galaxy structural components, which we derived in this study, with those found for external edge-on galaxies. The approach we used in this paper, when an integrated photometry is exploited to model the stellar distribution of a galaxy \citep[see e.g.][]{2010MNRAS.401..559M,2015ApJS..219....4S}, makes this comparison fair and direct. A careful examination of the galaxy scaling relations \citep[see e.g.][]{2010ApJ...722L.120F,1996ApJS..103..363C,2015MNRAS.451.2376M} will allow us to answer the question how typical our Galaxy is in comparison with other spiral galaxies.

\newpage
\section*{Acknowledgements}

This research is partly supported by the Russian Science Foundation grant no. 20-72-10052 for an analysis of the disc structure of the Galaxy and the influence of dust on its parameters. The authors also acknowledge financial support by the grant of the Russian Foundation for Basic Researches no. 19-02-00249 for an analysis of the B/PS-bulge structure using numerical simulations.

We thank the anonymous referee whose comments and valuable suggestions improved the paper.

This research makes use of data from the unWISE database \url{http://unwise.me/} based on products from the Wide-field Infrared Survey Explorer, which is a joint project of the University of California, Los Angeles, and the Jet Propulsion Laboratory/California Institute of Technology, and NEOWISE, which is a project of the Jet Propulsion Laboratory/California Institute of Technology. WISE and NEOWISE are funded by the National Aeronautics and Space Administration.


\section*{Data availability}
The data underlying this article will be shared on reasonable request to the corresponding author.



\bibliographystyle{mnras}
\bibliography{art} 





\bsp	
\label{lastpage}
\end{document}